\documentclass[twocolumn]{aastex6}

\received{08/10/2018}
\revised{11/27/2018}
\accepted{01/04/2019}

\usepackage{apjfonts}
\usepackage{aas_macros, xspace} 
\usepackage{comment}
\usepackage{hyperref}
\usepackage[flushleft]{threeparttable}
\usepackage{amsmath}
\usepackage{epstopdf}
\usepackage{lineno}

\newcommand{\eq}{\,=\,}

\def\ts     {\thinspace}
\def\kms    {\ts km\ts s$^{-1}$}

\def\msol   {$M_{\odot}$}

\def\lprime {K\,\kms\,pc$^2$}

\def\aco    {{\rm CO}($J$=1$\to$0)}

\def\bco    {{\rm CO}($J$=2$\to$1)}

\def\cco    {{\rm CO}($J$=3$\to$2)}

\def\eco    {{\rm CO}($J$=5$\to$4)}

\shorttitle{COLDz:\ CO Luminosity Function and Cold Gas History of the Universe}
\shortauthors{Riechers et al.}

\begin{document}

\title{COLDz:\ Shape of the CO Luminosity Function at High Redshift
  and the Cold Gas History of the Universe}

\author{Dominik A.\ Riechers\altaffilmark{1,2}}
\author{Riccardo Pavesi\altaffilmark{1}}
\author{Chelsea E.\ Sharon\altaffilmark{1,3,4}}
\author{Jacqueline A.\ Hodge\altaffilmark{5}}
\author{Roberto Decarli\altaffilmark{2,6}}
\author{\\ Fabian Walter\altaffilmark{2,7}}
\author{Christopher L.\ Carilli\altaffilmark{7,8}}
\author{Manuel Aravena\altaffilmark{9}}
\author{Elisabete da Cunha\altaffilmark{10}}
\author{Emanuele Daddi\altaffilmark{11}}
\author{\\ Mark Dickinson\altaffilmark{12}}
\author{Ian Smail\altaffilmark{13}}
\author{Peter L.\ Capak\altaffilmark{14}}
\author{Rob J.\ Ivison\altaffilmark{15,16}}
\author{Mark Sargent\altaffilmark{17}}
\author{Nicholas Z.\ Scoville\altaffilmark{18}}
\author{\\ Jeff Wagg\altaffilmark{19}}

\altaffiltext{1}{Cornell University, Space Sciences Building, Ithaca, NY 14853, USA}
\altaffiltext{2}{Max-Planck-Institut f\"ur Astronomie, K\"onigstuhl 17, D-69117 Heidelberg, Germany}
\altaffiltext{3}{Department of Physics \& Astronomy, McMaster University, 1280 Main Street West, Hamilton, ON L85-4M1, Canada}
\altaffiltext{4}{Yale-NUS College, \#01-220, 16 College Avenue West, Singapore 138527}
\altaffiltext{5}{Leiden Observatory, Leiden University, P.O. Box 9513, 2300 RA Leiden, The Netherlands}
\altaffiltext{6}{INAF - Osservatorio di Astrofisica e Scienza dello Spazio, via Gobetti 93/3, I-40129, Bologna, Italy}
\altaffiltext{7}{National Radio Astronomy Observatory, Pete V. Domenici Array Science Center, P.O. Box O, Socorro, NM 87801, USA}
\altaffiltext{8}{Cavendish Astrophysics Group, University of Cambridge, Cambridge, CB3 0HE, UK}
\altaffiltext{9}{N\'ucleo de Astronom\'ia, Facultad de Ingenier\'ia y Ciencias, Universidad Diego Portales, Av. Ej\'ercito 441, Santiago, Chile}
\altaffiltext{10}{Research School of Astronomy and Astrophysics, Australian National University, Canberra, ACT 2611, Australia}
\altaffiltext{11}{Laboratoire AIM, CEA/DSM-CNRS-Univ.\ Paris Diderot, Irfu/Service d'Astrophysique, CEA Saclay, Orme des Merisiers, F-91191 Gif-sur-Yvette cedex, France}
\altaffiltext{12}{National Optical Astronomy Observatory, 950 North Cherry Avenue, Tucson, AZ 85719, USA}
\altaffiltext{13}{Centre for Extragalactic Astronomy, Department of Physics, Durham University, South Road, Durham DH1 3LE, UK}
\altaffiltext{14}{Spitzer Science Center, California Institute of Technology, MC 220-6, 1200 East California Boulevard, Pasadena, CA 91125, USA}
\altaffiltext{15}{European Southern Observatory,
  Karl-Schwarzschild-Stra{\ss}e 2, D-85748 Garching, Germany}
\altaffiltext{16}{Institute for Astronomy, University of Edinburgh,
  Royal Observatory, Blackford Hill, Edinburgh EH9 3HJ, UK}
\altaffiltext{17}{Astronomy Centre, Department of Physics and Astronomy, University of Sussex, Brighton, BN1 9QH, UK}
\altaffiltext{18}{Astronomy Department, California Institute of Technology, MC 249-17, 1200 East California Boulevard, Pasadena, CA 91125, USA}
\altaffiltext{19}{SKA Organization, Lower Withington, Macclesfield, Cheshire SK11 9DL, UK}

 \email{riechers@cornell.edu}

\begin{abstract}

We report the first detailed measurement of the shape of the CO
luminosity function at high redshift, based on $>$320\,hr of the NSF's
Karl G.\ Jansky Very Large Array (VLA) observations over an area of
$\sim$60\,arcmin$^2$ taken as part of the CO Luminosity Density at
High Redshift (COLDz) survey. COLDz ``blindly'' selects galaxies based
on their cold gas content through \aco\ emission at $z$$\sim$2--3 and
\bco\ at $z$$\sim$5--7 down to a CO luminosity limit of log($L'_{\rm
  CO}$/\lprime )$\simeq$9.5. We find that the characteristic
luminosity and bright end of the CO luminosity function are
substantially higher than predicted by semi-analytical models, but
consistent with empirical estimates based on the infrared luminosity
function at $z$$\sim$2. We also present the currently most reliable
measurement of the cosmic density of cold gas in galaxies at early
epochs, i.e., the cold gas history of the universe, as determined over
a large cosmic volume of $\sim$375,000\,Mpc$^3$. Our measurements are
in agreement with an increase of the cold gas density from $z$$\sim$0
to $z$$\sim$2--3, followed by a possible decline towards
$z$$\sim$5--7. These findings are consistent with recent surveys based
on higher-$J$ CO line measurements, upon which COLDz improves in terms
of statistical uncertainties by probing $\sim$50--100 times larger
areas and in the reliability of total gas mass estimates by probing
the low-$J$ CO lines accessible to the VLA. Our results thus appear to
suggest that the cosmic star-formation rate density follows an
increased cold molecular gas content in galaxies towards its peak
about 10\,billion years ago, and that its decline towards the earliest
epochs is likely related to a lower overall amount of cold molecular
gas (as traced by CO) bound in galaxies towards the first billion
years after the Big Bang.

\end{abstract}


\section{Introduction} \label{sec:intro}

Our basic understanding of galaxy evolution and the build-up of
stellar mass in galaxies throughout the history of the universe is
founded in detailed measurements of the star-formation rate
density\footnote{Throughout this work, densities in star formation
  rate, stellar mass, or gas mass refer to cosmic densities (i.e., the
  amount of material in galaxies per unit co-moving cosmic volume)
  unless stated otherwise.} as a function of cosmic time (or
redshift), the ``star-formation history of the universe'', and
measurements of the stellar mass density in galaxies at different
cosmic epochs (see \citealt{md14} for a review). In-depth studies of
significant samples of high-redshift galaxies appear to indicate that
changes in this growth history at different epochs are largely driven
by the cold molecular gas properties of galaxies (e.g.,
\citealt{daddi10a,tacconi13,tacconi18,genzel15,scoville16}), and the
growth rate of dark matter halos (e.g.,
\citealt{genel10,bouche10,faucher11}). The cold gas constitutes the
fuel for star formation (see \citealt{cw13} for a review), such that a
higher gas content (e.g., driven by high gas accretion rates) or a
higher efficiency of converting gas into stars (e.g., driven by galaxy
mergers, or by ubiquitous shocks due to high gas flow rates) can lead
to increased star-formation activity, and thus, to a more rapid growth
of galaxies (e.g., \citealt{dave12}).

To better understand how the gas supply in galaxies moderates the
star-formation rate density in galaxies at early epochs, it is
desirable to complement targeted studies with an integrated
measurement of the cold molecular gas density in galaxies at the same
epochs, i.e., the ``cold gas history of the universe''. Surveys of
cold gas ideally target rotational lines of CO, the most common tracer
of the molecular gas mass in galaxies, to measure the CO luminosity
function at different cosmic epochs in a ``molecular deep field''
study. The distribution mean of the CO luminosity function then
provides a reliable measurement of the cold molecular gas density at a
given redshift (\citealt{cw13}; see, e.g., \citealt{scoville17} for an
alternative approach). The first such efforts have recently been
carried out in the {\em Hubble} Deep Field North (HDF-N) and the {\em
  Hubble} Ultra Deep Field (H-UDF) with the IRAM Plateau de Bure
Interferometer (PdBI) and the Atacama Large Millimeter/submillimeter
Array (ALMA; the ASPECS-Pilot program) at 3\,mm and 1\,mm wavelengths,
covering fields $\sim$0.5 and $\sim$1\,arcmin$^2$ in size,
respectively (see \citealt{decarli14,walter16}, and references
therein). At $z$$\sim$2--3, near the peak of the cosmic star-formation
rate density $\sim$10\,billion years ago, these studies cover
\cco\ and higher-$J$ lines. At $z$=5--7, i.e., in the first billion
years after the Big Bang, these surveys cover \eco\ and higher-$J$
lines.

The most faithful tracer of total cold gas mass are low-$J$ CO lines,
in particular, \aco\ (see, e.g.,
\citealt{riechers06c,riechers11e,riechers11c,ivison11,aravena12,aravena14,daddi15,bolatto15,sharon16,saintonge17,harrington18}),
for which the $\alpha_{\rm CO}$=$M_{\rm H_2}$/$L'_{\rm CO}$ conversion
factor from CO luminosity ($L'_{\rm CO}$, in units of \lprime ) to
H$_2$ gas mass ($M_{\rm H_2}$, in units of \msol ) has been calibrated
locally (see \citealt{bolatto13} for a review), and for which no
assumptions about gas excitation are required to derive the total CO
luminosity. To complement the initial ``molecular deep field'' studies
through improved statistical uncertainties measured over larger cosmic
volumes and reduced calibration uncertainties due to gas excitation,
we have carried out the VLA COLDz survey,\footnote{See {\tt
    coldz.astro.cornell.edu} for additional information.} ``blindly''
selecting galaxies through their cold gas content in the \aco\ line at
$z$$\sim$2--3, and in \bco\ at $z$$\sim$5--7, over a
$\sim$60\,arcmin$^2$ region.

The detailed survey parameters, line search and statistical
techniques, a catalog of line candidates, and an overview of
accompanying papers are presented in the COLDz survey reference paper
(\citealt{pavesi18}; hereafter: Paper I). This work focuses on the CO
luminosity function measurements to result from the survey data, and
the implied constraints on the evolution of the cosmic cold gas
density in galaxies as a function of redshift.

Section 2 provides a brief description of the data. Section 3
summarizes the selection of CO line candidates and the statistical
methods used to characterize the survey parameters, before describing
the CO luminosity function and cold gas density measurements. Section
4 provides a discussion of the results in the context of previous
surveys and model predictions. Section 5 provides the main conclusions
based on our measurements and analysis. We use a concordance, flat
$\Lambda$CDM cosmology throughout, with
$H_0$\eq69.6\,\kms\,Mpc$^{-1}$, $\Omega_{\rm M}$\eq0.286, and
$\Omega_{\Lambda}$\eq0.714 (\citealt{bennett14}).

\section{Data} \label{sec:data}

The VLA was used to ``blindly'' observe redshifted \aco\ and
\bco\ line emission (rest-frame frequencies:\ $\nu_{\rm
  rest}$=115.2712 and 230.5380\,GHz) at $z$$\sim$2.0--2.8 and
$z$=4.9--6.7, respectively (VLA program IDs 13A-398 and 14A-214;
PI:\ Riechers), covering a region corresponding to a total co-moving
survey volume of $\sim$375,000\,Mpc$^3$ in both lines combined (see
Table~\ref{t1} for details). Observations covered a 7-pointing mosaic
in the COSMOS field (center position:\ J2000 10:00:20.7; +02:35:17.0),
and a 57-pointing mosaic in the GOODS-North field (center
position:\ J2000 12:36:59.0; +62:13:43.5), providing total survey
areas of 8.9 and 50.9\,arcmin$^2$ at 31 and 30\,GHz,
respectively.\footnote{The mosaicked images made from individual
  pointings were trimmed at the outer edges at the 30\% level of the
  peak sensitivity in each spectral channel.}

\begin{figure}
\epsscale{1.15}
\plotone{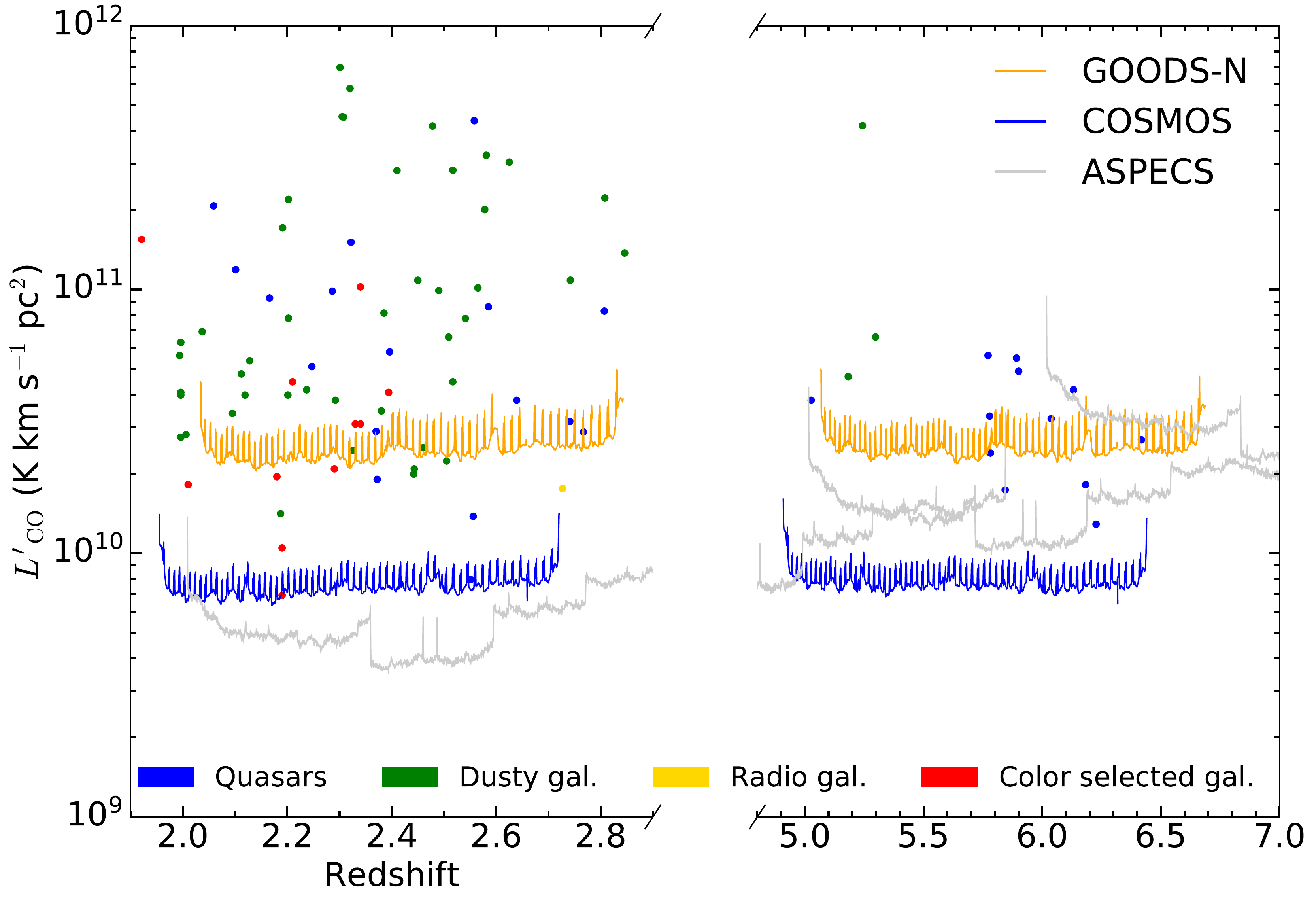}
\vspace{-2mm}

\caption{Representative line luminosity detection sensitivity limits
  as a function of redshift reached by our observations in the COSMOS
  and GOODS-North fields, when adopting 5 times the rms noise at a
  line FHWM of 200\,\kms\ (see Paper I for variations between
  individual pointings). The corresponding sensitivity limits of the
  ALMA ASPECS-Pilot survey in the CO $J$=3$\to$2 to 7$\to$6 lines in
  the same redshift ranges are shown for comparison
  (\citealt{walter16}). ASPECS limits are scaled to \aco\ line
  luminosity limits using the same, representative line excitation
  correction factors adopted by \cite{decarli16a} based on
  \cite{daddi15} up to \eco. For the higher-$J$ lines, we assume an
  intermediate excitation based on Fig.~10 of \cite{daddi15},
  corresponding to brightness temperature ratios of
  $r_{65}$$\simeq$0.66 and $r_{75}$$\simeq$0.29 relative to the CO
  $J$=5$\to$4 line, respectively. For reference, colored points show
  all previous $z$$>$1 CO detections as compiled by \cite{cw13},
  incorporating updates by \cite{sharon16}. Colors indicate different
  galaxy types (``dusty galaxies'' includes submillimeter galaxies,
  extremely red objects, and 24\,$\mu$m-selected galaxies, and
  ``color-selected galaxies'' includes Lyman-break, BzK, and BMBX
  galaxies, respectively).
  \label{f1}}
\end{figure}

Observations in COSMOS and GOODS-North covered the 30.969 to 39.033
and 29.981 to 38.001\,GHz frequency ranges in a single tuning with
$\sim$8\,GHz bandwidth (dual polarization) each, respectively, using
the Ka band receivers and the 3-bit samplers at a spectral resolution
of 2\,MHz (17\,\kms\ at 35\,GHz). Observations were carried out for a
total of 324\,hr between 2013 January 26 and 2015 December 18 under
good to excellent weather conditions in the D and DnC array
configurations, and in the D$\to$DnC and DnC$\to$C re-configurations,
yielding approximately 93 and 122\,hr on source across all
configurations and pointings in COSMOS and GOODS-North,
respectively. This yields characteristic synthesized beam sizes of
$\sim$3$''$ when imaging the data with natural baseline weighting
after data calibration using the {\sc CASA} package, and approximately
3 times better point source sensitivity in the smaller COSMOS
mosaic. The corresponding CO luminosity limits across the redshift
range are shown in Fig.~\ref{f1}. More details on the observations and
data reduction are given in Paper I.

\begin{table}
\caption{Lines, redshift ranges, and volumes covered by the COLDz survey (see Fig.~\ref{f1} for luminosity limits across the survey volume).\label{t1}}
\centering
\begin{tabular}{ c c c c c c }
\hline
  Transition & $\nu_{\rm rest}$ & $z_{\rm min}$ & $z_{\rm max}$ & $\langle z \rangle$ & Volume \\
   & [GHz] & & & & [$\rm{Mpc}^3$] \\
   \hline
   \multicolumn{6}{c}{COSMOS ``Deep'' Field ($\sim$9\,arcmin$^2$; $\sim$31--39\,GHz)}\\
   \aco & 115.271 & 1.953 & 2.723 & 2.354 & 20,189\\
   \bco & 230.538 & 4.906 & 6.445 & 5.684 & 30,398\\
  \hline
\multicolumn{6}{c}{GOODS-North ``Wide'' Field ($\sim$51\,arcmin$^2$; $\sim$30--38\,GHz)}\\
   \aco & 115.271 & 2.032 & 2.847 & 2.443 & 131,042\\
   \bco & 230.538 & 5.064 & 6.695 & 5.861 & 193,286\\
  \hline
\end{tabular}
 \tablecomments{The co-moving volume is calculated to the edges of the
   mosaics, and does not account for varying sensitivity across the
   mosaics, which is accounted for by the completeness correction.}
\end{table}

\section{Results and Analysis}

\subsection{CO Line Candidates}

Based on our matched filtering algorithm using three-dimensional
spatial/spectral templates, we find 57 candidate \aco\ and \bco\ line
emitters in our survey volume down to signal-to-noise ratio (SNR)
limits of 5.25 and 5.50 in the COSMOS (26 candidates) and GOODS-North
(31 candidates) fields, respectively (Paper I). These SNR limits are
chosen to provide comparable ratios of line candidates with positive
fluxes over noise spikes with negative fluxes at matching SNR between
both fields. This misses at least one independently confirmed
\bco\ emitter, HDF\,850.1 at $z$=5.18 in the GOODS-North field
(\citealt{walter12}), which has a SNR of 5.33. Including this source,
7 out of the 58 candidates are presently independently confirmed to be
real \aco\ (three sources in COSMOS, one in
GOODS-North)\footnote{Sources are COLDz.COS.1 to 3 and COLDz.GN.3
  (GN19) in Paper I.} or \bco\ line emitters (one in COSMOS, two in
GOODS-North)\footnote{Sources are COLDz.COS.0 (AzTEC-3), and
  COLDz.GN.0 (GN10) and 31 (HDF\,850.1) in Paper I.} through the
detection of additional CO lines (see, e.g., Paper I;
\citealt{riechers10a,riechers11c,riechers14b}; \citealt{walter12}).
To remain robust against individual, potentially spurious candidates,
all other line candidates are used only in the statistical analysis of
the survey data in a probabilistic manner until independent
confirmation is obtained. All candidates except the three
independently-confirmed \bco\ emitters are considered to be
\aco\ emitters unless stated otherwise.\footnote{This initial
  assumption is based on the expectation of a significantly lower
  space density of bright CO-emitting galaxies at $z$$\gtrsim$5
  compared to $z$$<$3 in our current understanding. Alternative
  scenarios are also discussed below.}

\subsection{Statistical Methods}

A detailed description of the statistical properties of the candidate
CO line emitter sample is given in Paper I. We here briefly summarize
the main elements of the methods as relevant to the construction of
the CO luminosity function. The main purpose of the statistical
analysis is to determine the probability of each line candidate to be
real, and the completeness of the line search as a function of line
luminosity, spatial size, and velocity width. Furthermore, it is
necessary to evaluate the probability function of the actual versus
measured line luminosity.

\subsubsection{Reliability Analysis:\ Purity Estimates}

The reliability (or purity/fidelity) of each candidate CO emitter is
given by its probability of corresponding to a real line source. The
reliability analysis in this work employs Bayesian techniques based on
the assumption of symmetry of the noise distribution in the data cubes
around zero flux, identifying real signal as positive ``excess'' flux
at a given SNR evaluated relative to the noise distribution as traced
by negative flux features found with the same extraction methods. The
posterior probability distributions for the rates of real sources and
noise spikes (which provide estimates of the purity) are obtained by
modeling the occurence rates of real sources and noise spikes as an
inhomogenous Poisson process with different rate models in the SNR
distribution. Purities are calculated from the posterior probability
distributions as a function of the model parameters, which are sampled
using a Markov Chain Monte Carlo (MCMC) technique (emcee;
\citealt{foreman13}). The Poisson rate of the noise distribution is
modeled as the tail of a Gaussian in SNR, based on negative flux
features at SNR $\geq$4 in the data. The occurence rate of noise
features is then measured by maximizing the likelihood of the noise
model based on all negative features down to the above SNR limit. The
real source rate is parameterized as a shallow power law increasing
towards lower SNR values (based on the conservative assumption that
faint sources are more common), normalized at a SNR of 6, with
uniform, non-constraining priors on the slope and normalization. Given
the simple parameterization as a slowly varying source rate, we
include all candidates in the limited 5$<$SNR$<$9 range where most
candidates are found in the analysis, after confirming that candidates
with SNR$>$9 (which represent rare sources in our survey) are always
assigned a purity of 100\%. COLDz.GN.3 (GN19) is the only
independently confirmed \aco\ emitter with a SNR$<$9, and thus, the
only secure source considered in the statistical analysis that would
be assigned a purity of $<$100\% by this method alone. Taking into
account the additional information that confirms this source to be
real, its purity however is known to be 100\% in practice. Thus, the
latter, preferred value is adopted in the subsequent
analysis. Following this change, all independently confirmed
\aco\ emitters have a purity of 100\%$\pm$0\% (see Paper I, Appendix
F.1, for further details).

\subsubsection{Artificial Source Analysis:\ Completeness and Flux Corrections}

The line search and flux extraction methods employed by a ``blind''
survey, in combination with the observational parameters of the data
cubes, determine the completeness of a survey. The choice of methods
however may introduce biases in the measurements. A probabilistic
analysis of artificial sources of varying fluxes and three-dimensional
sizes (i.e., spatial extent and linewidth) injected into the data
cubes\footnote{The data cubes are used to represent the noise
  properties, since the vast majority of resolution elements are void
  of signal.} and then re-extracted using the same methods as for real
candidates is employed here to estimate survey completeness, and to
account for potential biases. A statistical comparison of injected to
extracted artificial source properties is used to correct the CO
luminosity function for the completeness of our line search. This
method is also used to correct the measured source fluxes feeding into
the luminosity function in a probabilistic manner by estimating biases
in the flux extraction procedure\footnote{Flux extraction relies on
  fitting Gaussian line profiles to aperture spectra extracted over a
  FWHM diameter determined from two-dimensional Gaussian size fitting
  to velocity-integrated line maps, and thus, is sensitive to the
  fitted size.} and uncertainties in the flux recovery. Using Bayes
theorem, the size, line velocity width, and flux probability
distributions found from the artificial source analysis are used in
combination with a conservative prior on the fraction of resolved
sources to find the probabilistic relationships between measured and
real source sizes, line velocity widths, and fluxes. The completeness
of the detection process is determined by measuring the fraction of
the injected sources that are detected for a given set of source
parameters (i.e., line flux, velocity width, and spatial size) as a
function of SNR. This provides an estimate for the fraction of the
objects at a given intrinsic line luminosity that are recovered at a
given SNR threshold, and by accounting for variations of the
sensitivity as a function of position and frequency in the mosaics,
over what fraction of the survey volume they can be recovered. In the
construction of the luminosity function, the probability-weighted
completeness is determined by assuming for each luminosity bin that
the frequency and line luminosity distributions are uniform within the
bin, and no intrinsic spatial size and line velocity width
distributions are assumed (i.e., the measured values of the candidates
are adopted to determine completeness). Mean values, rather than ``per
source'' completeness values of individual candidates, are adopted for
each bin of the luminosity function, to minimize biases (see Paper I,
Appendix F.3, for further details).

\subsubsection{Construction of the CO Luminosity Function}

The CO luminosity function is assembled by weighting the contribution
of each line candidate to a given luminosity bin by its purity and, in
a statistical manner (i.e., per luminosity bin, not per source),
inversely by its completeness, using the total co-moving cosmic volume
covered by the survey (which corresponds to an effective volume
$V_{\rm max}$ for each candidate galaxy due to the completeness
corrections and spatial variations of the survey
sensitivity). Systematic uncertainties are estimated by calculating
the luminosity function with random realizations of the flux and
purity estimates (within their statistically well-defined
distributions), and with different luminosity bin widths and boundary
conditions.

In each luminosity bin, the completeness is determined by averaging
over 1000 random realizations (i.e., using a uniform prior) for each
combination of spatial size and velocity width covered by the
artificial sources, using randomly sampled redshifts to calculate the
line fluxes as a basis for the completeness values. We then apply
these values as a function of spatial size and velocity width to the
probability distribution in these parameters for each candidate,
effectively using the parameters as weights to the completeness.

Systematic uncertainties are taken into account by calculating 10,000
Monte Carlo realizations of the luminosity function for every
luminosity bin width and center (sampled in intervals of 0.1\,dex),
varying purity values and flux assignments (which are drawn from
log-normal distributions for different spatial sizes; see Paper I) for
each candidate independently (i.e., allowing them to shift between
adjacent luminosity bins) to simulate the uncertainty in their
intrinsic fluxes. In the final analysis, luminosity bin widths of
0.5\,dex are adopted. A conservative 20\% uncertainty is added to
measured fluxes to account for the uncertainty in flux calibration of
the data.\footnote{This value is higher than the nominal precision of
  absolute flux calibration at the VLA, to account for the fact that
  some observing runs did not contain one of the ``standard'' primary
  flux calibrators (see Paper I for details). This conservative choice
  has only a minor impact on our results.} From the 10,000
realizations, median values and the scatter around the median are
calculated for each luminosity bin. The scatter is expressed as the
95$^{\rm th}$ and 5$^{\rm th}$ percentiles for the upper and lower
bounds, respectively. Statistical Poisson uncertainties are calculated
for each luminosity bin as the relative uncertainty of 1/$\sqrt{N}$,
where $N$ corresponds to the number of candidates in the bin (see
Paper I, Appendix F.4, for further details).

\begin{figure*}
\epsscale{1.15}
\plotone{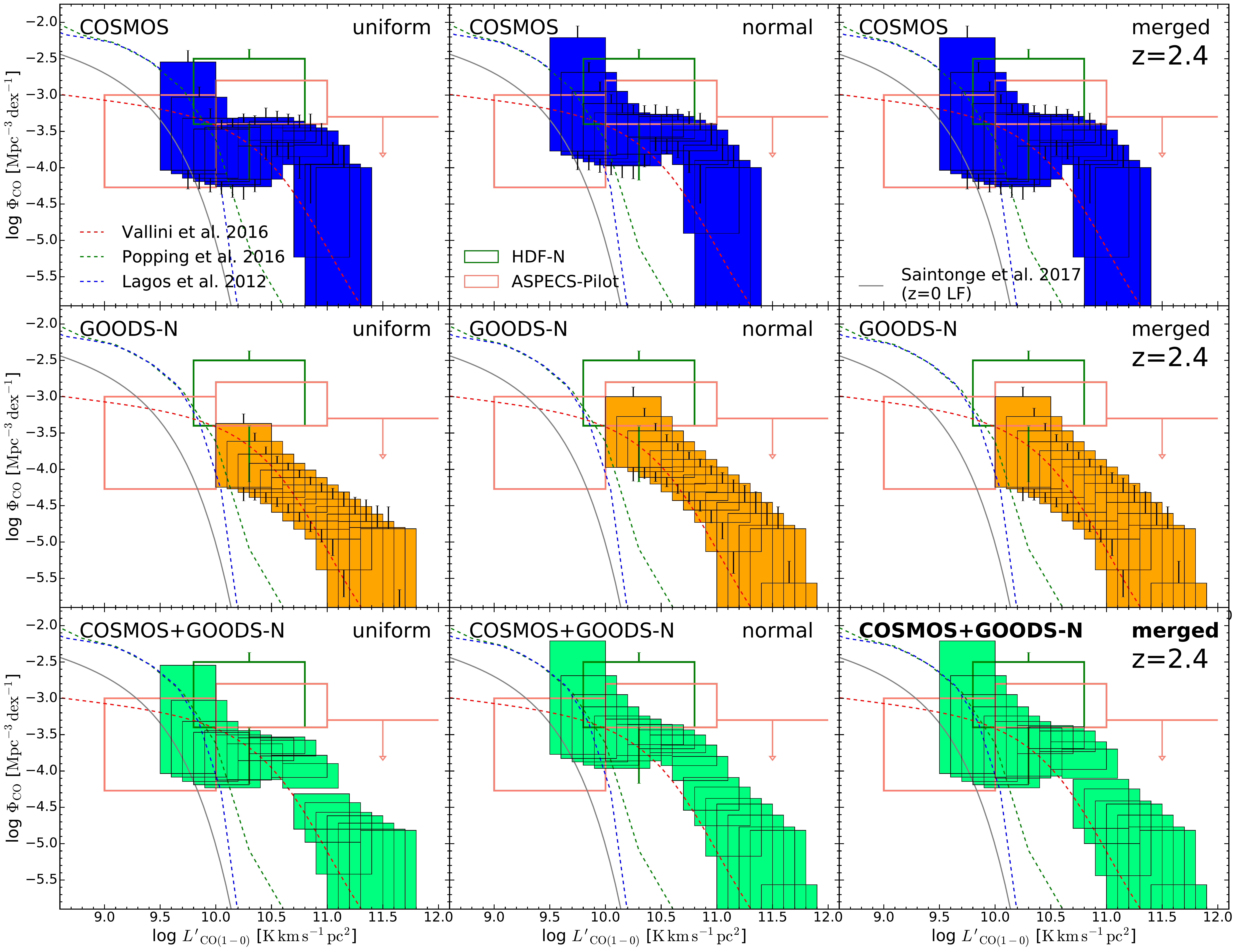}
\vspace{-2mm}

\caption{VLA COLDz \aco\ luminosity function at
  $\langle$$z$$\rangle$=2.35 and 2.44 in the COSMOS ({\em top}) and
  GOODS-North ({\em middle}) fields (shaded boxes), respectively, and
  the combination of both fields, weighted by the statistical
  uncertainties in each field ({\em bottom}), showing the consistency
  between methods and fields. The {\em left} panels show the
  constraints obtained when conservatively assuming a uniform
  distribution between zero and the most likely values for the
  purities. The {\em middle column} panels show the constraints when
  assuming the most likely values for the purities and assigning 100\%
  uncertainty to these values, truncated at zero and 100\%. The {\em
    right} panels show the composite uncertainties merged from both
  methods, obtained by assuming the lowest and highest values covered
  by their respective uncertainty ranges in each bin. Bins have a
  width of 0.5\,dex in $L'_{\rm CO}$, and step through the covered
  luminosity range in steps of 0.1\,dex. As such, individual bins are
  not statistically independent. Error bars on the boxes indicate
  Poissonian uncertainties in each bin. Empty green boxes are the
  constraints on the \cco\ luminosity function at
  $\langle$$z$$\rangle$=2.75 from the PdBI HDF-N survey
  (\citealt{walter14}). Empty orange boxes are the \cco\ constraints
  at $\langle$$z$$\rangle$=2.61 from the ALMA ASPECS-Pilot survey
  (\citealt{decarli16a}). A constant \cco/\aco\ brightness temperature
  ratio of $r_{31}$=0.42 has been applied to correct the
  \cco\ luminosities to \aco\ luminosities for both these surveys. The
  gray line shows the $z$=0 luminosity function for comparison
  (updated from \citealt{saintonge17}; A.\ Saintonge, private
  communication). Dashed lines are semi-analytical and empirical
  model predictions (\citealt{lagos12,popping16,vallini16}). All
  except the COLDz data are the same in all panels.
  \label{f2}}
\end{figure*}

To further account for systematic uncertainties not fully captured by
our statistical treatment, purities are utilized using two different
methods. The first method (termed ``normal'' hereafter; Fig.~\ref{f2},
middle) draws the purity values for the MCMC sampling used to estimate
the allowed range for the luminosity function as random numbers with a
normal distribution around the computed values, with a standard
deviation of the values themselves, truncated at zero and 100\%
purity. This method is motivated by the modest number of line
candidates in excess of the tail end of the noise distribution (see
Paper I), which limits the precision of more direct methods of
measuring the uncertainties. The second method (``uniform'';
Fig.~\ref{f2}, left) more conservatively treats the purities as upper
limits, and draws the purity values from a uniform distribution
between zero and the computed values. This method is designed to
account for the finding that a significant fraction of moderate
significance line candidates do not show unambiguous multiwavelength
counterparts (see Paper I), motivating a more conservative treatment
of the purity prior. The results from these implementations are
consistent within the uncertainties. In the following, we thus
conservatively adopt a combination of both methods by assuming the
outermost upper and lower bounds of the uncertainties between the two
methods (Fig.~\ref{f2}, right).

Given the limited survey statistics due to the moderate number of line
candidates, luminosity function constraints are displayed in bins that
are not statistically independent throughout, but which instead sample
the luminosity function in luminosity bins of 0.5\,dex width, in steps
of 0.1\,dex. Independent bins thus are recovered by only considering
every fifth bin (see Table~\ref{ta1}). Since candidates are primarily
taken into account in a probabilistic manner instead of on a
per-candidate basis, this choice of partially redundant sampling does
reveal additional information on the shape of the luminosity function,
and shows trends more clearly than broader, more sparsely sampled
bins.

\begin{figure*}
\epsscale{1.15}
\plotone{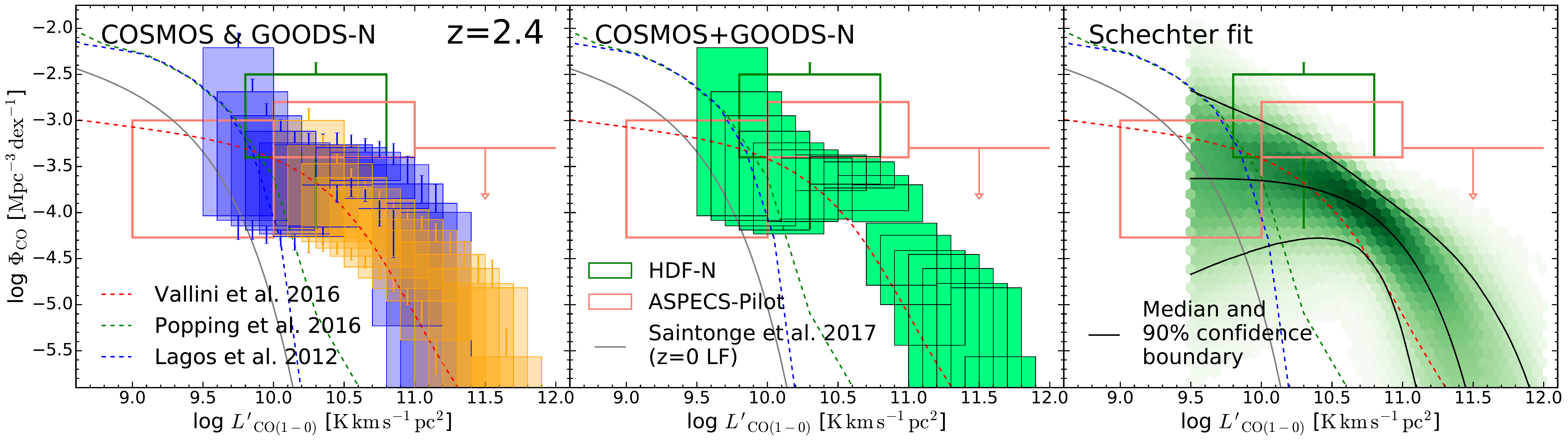}
\vspace{-2mm}

\caption{Comparison between the two COLDz survey fields and combined
  constraints on the \aco\ luminosity function. {\em Left:} Same as
  Fig.~\ref{f2}, top and middle right, but showing the COSMOS and
  GOODS-North constraints overlaid with each other. {\em Middle:} Same
  as Fig.~\ref{f2}, bottom right, combining the measurement from both
  fields, for comparison.  {\em Right:} Density of Schechter function
  fits to the combined data, as distributed according to the fit
  parameter distributions as obtained with the ABC method (shaded
  region). Darker colors represent higher probabilities. For
  reference, solid black lines show the median and 90\% confidence
  boundary of the implied luminosity function distributions (see
  Tab.~\ref{t2} for corresponding Schechter parameters).  All except
  the COLDz data are the same as in Fig.~\ref{f2} in all
  panels. \label{f3}}
\end{figure*}

\begin{figure}
\epsscale{1.15} \plotone{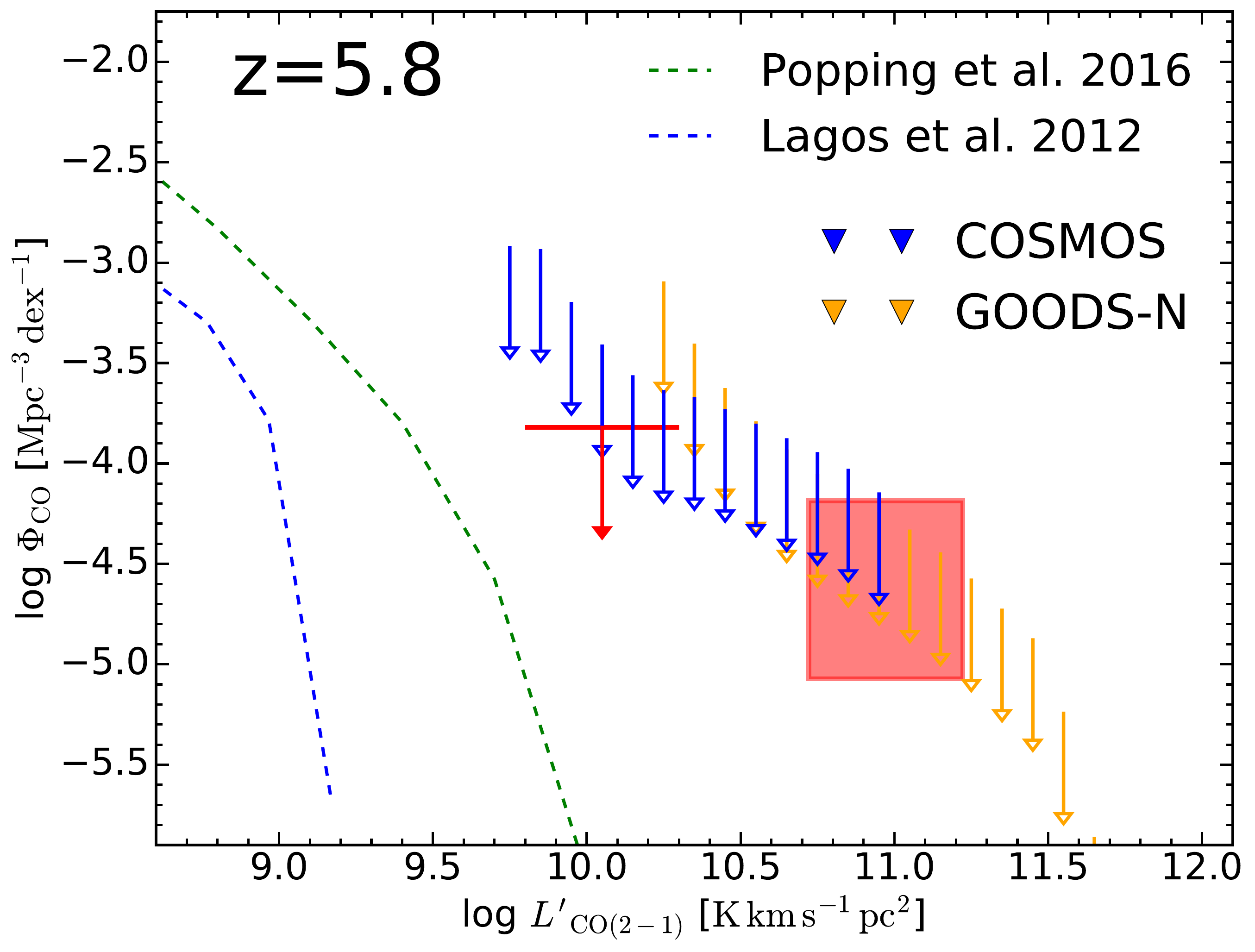}
\vspace{-2mm}

\caption{VLA COLDz \bco\ luminosity function at
  $\langle$$z$$\rangle$=5.68 and 5.86 in the COSMOS (blue and red
  arrows) and GOODS-North (orange arrows and red boxes) fields,
  respectively. The blue and orange arrows show upper limits under the
  unlikely assumption that any candidates not independently confirmed
  to be \aco\ emission would correspond to \bco\ emission, for the
  same binning in $L'_{\rm CO}$ as in Fig.~\ref{f2}. Red arrows and
  boxes consider only independently confirmed \bco\ candidates. The
  dashed lines show model predictions. \label{f4}}
\end{figure}

\begin{table}
\caption{Schechter function fit parameter constraints to the \aco\ luminosity function at $z$=1.95--2.85 from COLDz. \label{t2}}
\centering
\begin{tabular}{ c c c c }
\hline
  Parameter & 5$^{\rm th}$ percentile & 50$^{\rm th}$ percentile & 95$^{\rm th}$ percentile\\
   \hline
  $\log L'^*_{\rm CO}$   & 10.22  & 10.70  & 11.33\\
  $\log \Phi^*_{\rm CO}$ & --4.66 & --3.87 & --3.20\\
  $\alpha$              &--0.78 & 0.08  & 0.99\\
  \hline
\end{tabular}
\tablecomments{The CO luminosity function is defined as 
  $\log \Phi_{\rm CO}=\log \Phi^*_{\rm CO}+  \alpha\,(\log L'_{\rm CO}-\log L'^*_{\rm CO})-L'_{\rm CO}/(L'^*_{\rm CO}\,{\rm ln\,10})+\log({\rm ln\,10})$. $L'$ is given in units of K\,\kms\,pc$^2$. $\Phi$ is given in units of Mpc$^{-3}$\,dex$^{-1}$.}
\end{table}

\subsection{COLDz CO Luminosity Function}

\subsubsection{\aco\ Luminosity Function}

The estimates of the \aco\ luminosity function, which include all
candidates except independently confirmed \bco\ emitters,\footnote{The
  level of the CO luminosity function is dominantly determined by
  independently-confirmed sources, such that unconfirmed candidates
  mainly contribute to the size of the uncertainty ranges. For
  reference, in the COSMOS field, 1, 2, or 3 secure detections in a
  bin correspond to log\,$\Phi_{\rm CO}$ [Mpc$^{-3}$\,dex$^{-1}$]
  =--4.00, --3.70, or --3.53, respectively. In the GOODS-North field,
  the same number of detections correspond to values of --4.82,
  --4.52, and --4.34, respectively.} are consistent\footnote{There are
  some apparent variations between the two fields within the
  uncertainties, e.g., around the $\log(L'_{\rm CO}$/\lprime )
  =10.5--11.0 bin, which are likely a reflection of cosmic variance.}
between both survey fields (Fig.~\ref{f3}, left). We thus decided to
merge the constraints from both fields through a weighted average in
each bin (Fig.~\ref{f2}, bottom, and Fig.~\ref{f3}, middle).

The data reveal the shape of the \aco\ luminosity function at
$z$$\sim$2.4, which resembles that of a Schechter function. While not
a unique solution given current observational constraints, we obtain
an estimate of the allowed range of Schechter parameters by fitting
the characteristic parameters $L'^*_{\rm CO}$ and $\Phi^*_{\rm CO}$
and the power-law slope $\alpha$ to the data (Fig.~\ref{f3}, right and
Table \ref{t2}). We adopt the Approximate Bayesian Computation (ABC)
method (see, e.g., \citealt{cameron12,weyant13,ishida15}, and
references therein) to derive posterior distributions of the Schechter
parameters, in order to account for all selection effects affecting
our measurement without having to specify an explicit equivalent
likelihood function.

We first assume uniform, unconstraining priors on the Schechter
parameters (i.e., $L'^*_{\rm CO}$, $\Phi^*_{\rm CO}$, and $\alpha$) to
describe the intrinsic distribution of the CO luminosity
density. Following the ABC method, we then iteratively sample from
these priors, randomly sampling galaxies with randomly assigned line
widths and spatial sizes (according to the same prior distributions as
assumed before) from the assumed distributions. Here, the number of
galaxies is assumed to follow a Poisson distribution around the
mean. We then correct for completeness and flux recovery in our line
search to determine a mock observed sample of galaxies. We then
compare this mock sample to observations by only including line
candidates with probability equal to their purity, where purities are
a random variable determined according to the ``normal'' and
``uniform'' methods described in Sect. 3.2.3.

The decision criterion for a mock sample to provide a sufficient match
to the observed sample, and thus, to retain an initial set of
Schechter parameters for the posterior distribution, is to result in
the same number of sources as observed, with line fluxes matching to
within the 20\% flux calibration uncertainty. For each of the survey
fields, the process is repeated until sufficient accepted samples are
generated to accurately define the posterior distribution. The same
procedure is repeated for both survey fields combined, requiring that
the resulting data set can be represented by a single, common CO
luminosity function.\footnote{For this last run, the number of mock
  sources was allowed to differ by one from the observed sample, which
  we expect to have a minor impact on the result.} In a final step, we
merge the results from both purity methods by giving each method equal
probability. The resulting model parameter posterior distributions
are shown in Fig.~\ref{fb1}.

As shown in Figs.~\ref{f3} and \ref{fb1}, uncertainties are dominated
by the faint-end slope below the ``knee'', given the sensitivity of
the survey. On the other hand, $L'^*_{\rm CO}$, $\Phi^*_{\rm CO}$ are
constrained fairly reliably by the data, with a reasonable agreement
within the uncertainties between the two survey fields.

\subsubsection{\bco\ Luminosity Function}

For estimates of the \bco\ luminosity function, we followed two
approaches. The first approach excludes all candidates used to
construct the \aco\ luminosity function from our search, only leaving
confirmed \bco\ sources and upper limits as available constraints (see
Paper I). We further exclude one of the \bco\ sources, AzTEC-3 in the
COSMOS field (\citealt{riechers10a,riechers14b,capak11}), because the
field was chosen to include this source as a bright ``calibrator'' for
the line search methods. Including this source thus would likely bias
the measurement in its luminosity bin towards high values, under the
reasonable assumption that a random $\sim$9\,arcmin$^2$ field would be
unlikely to include a $z$$>$5 source as luminous as AzTEC-3. As lower
and upper bounds on the uncertainties, the 5$^{\rm th}$ and 95$^{\rm
  th}$ percentiles of the Bayesian posterior for inferring a Poisson
rate are adopted in GOODS-North (where sources are detected), and the
68$^{\rm th}$ percentile is adopted as an upper limit for the COSMOS
field (where no secure CO $J$=2$\to$1 detections remain after the
exclusion of AzTEC-3). The second approach assumes that all
\aco\ candidates that are not independently confirmed could
potentially be \bco\ emitters at higher redshifts. Although unlikely,
this provides shallower but more detailed upper limits on the
\bco\ luminosity function in smaller luminosity bins. Estimates are
weighted by purities and corrected for completeness, using the
``normal'' purity method, adopting the upper, 90$^{\rm th}$ percentile
bounds as upper limits. The results from both methods are consistent
with each other, and are shown together in Figure \ref{f4}.

\begin{figure*}[th!]
\epsscale{0.58}
\plotone{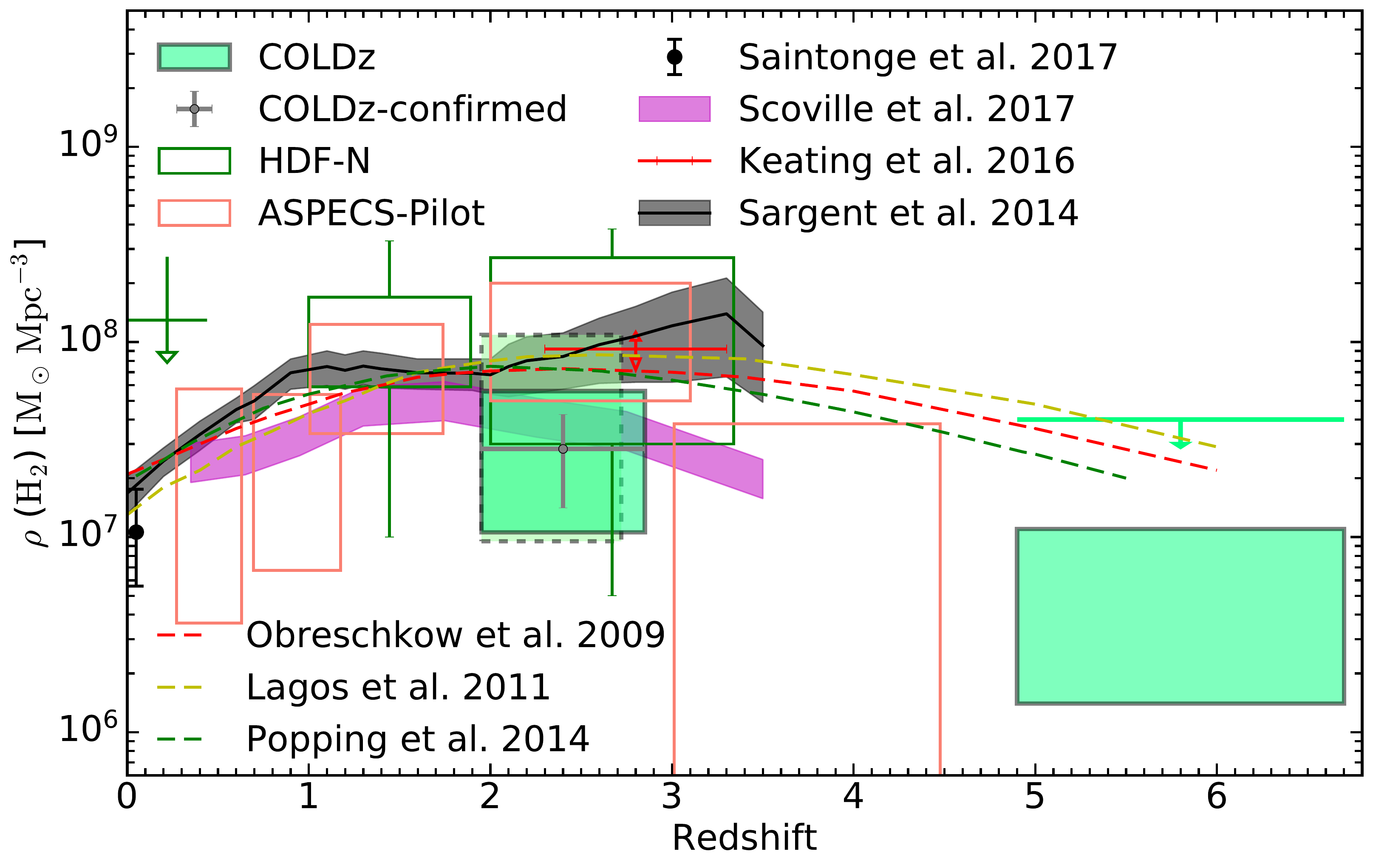}
\plotone{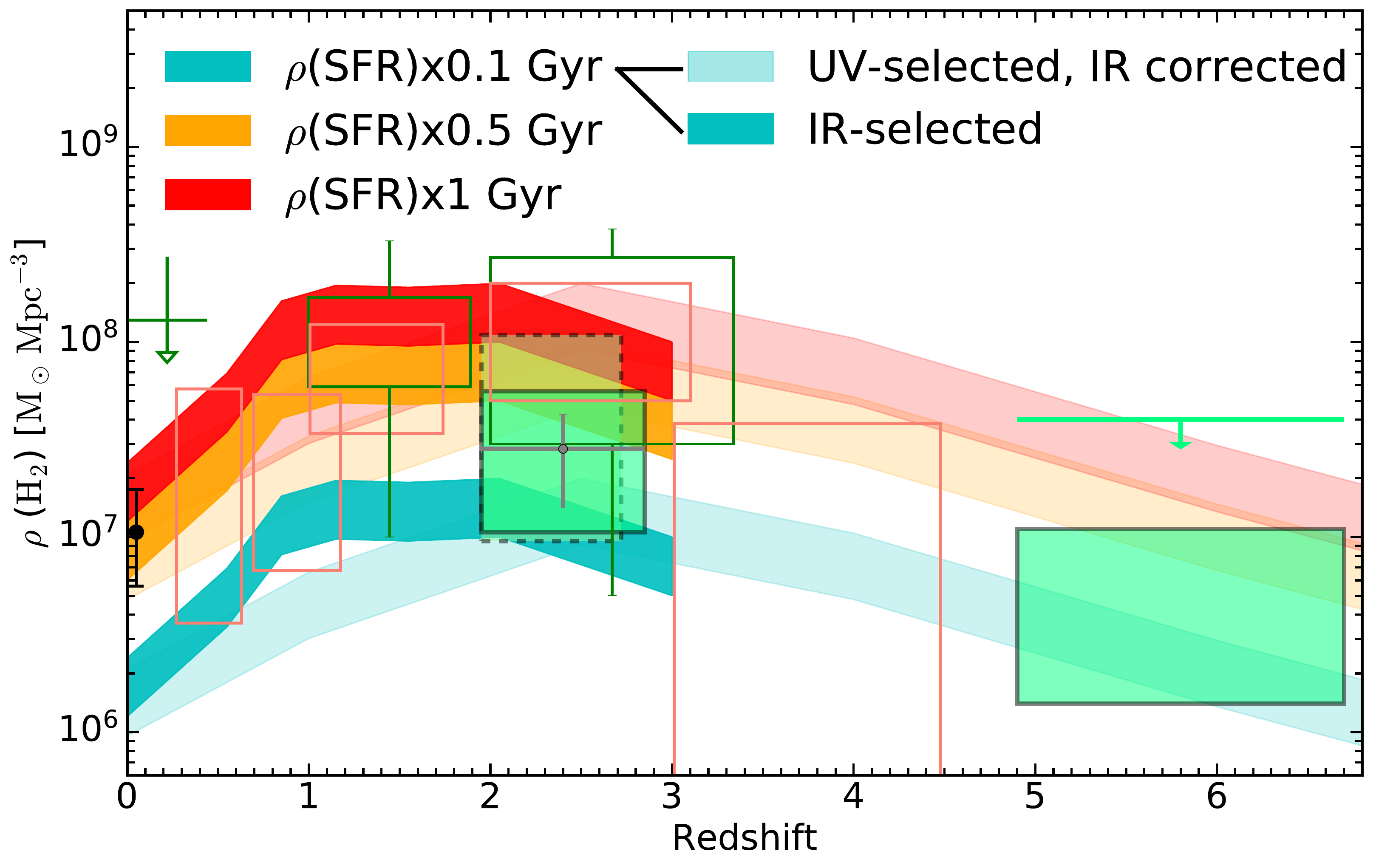}
\vspace{-5.5mm}

\caption{VLA COLDz measurements of the cold gas history of the
  universe (green boxes), i.e., the co-moving cosmic mass density of
  cold molecular gas as a function of redshift, showing that the gas
  density evolves. Vertical sizes indicate the uncertainties in each
  bin. In the $z$$\simeq$2--3 bin, the smaller solid box shows the
  constraints from both fields combined, and the larger dashed box
  shows the constraints from the COSMOS field only (both after merging
  the two purity methods), as an illustration of the impact of
  field-to-field variations. Assumptions for the measurement and
  uncertainties in the $z$=4.90--6.70 bin are the same as in
  Fig.~\ref{f4}. For reference, the gray point shows the measurement
  obtained when only including independently confirmed candidates,
  which is fully consistent with the measurement obtained from the
  complete statistical analysis. Empty green and orange boxes show the
  constraints from the same surveys as in Fig.~\ref{f2}, where
  different boxes correspond to estimates obtained in different CO
  transitions (\citealt{walter14,decarli16a}). The black point shows
  constraints at $z$=0 (\citealt{saintonge17}). {\em Left:} Dashed
  lines show model predictions
  (\citealt{obreschkow09,lagos11,popping14b,popping14a}). The gray
  shaded range shows empirical predictions based on an inversion of
  the $M_{\rm gas}$--SFR relation (e.g.,
  \citealt{sargent12,sargent14}; scaled to $\alpha_{\rm
    CO}$=3.6\,\msol\,(K\,\kms\,pc$^{2}$)$^{-1}$ from its original
  effective value of
  $\sim$4.4\,\msol\,(K\,\kms\,pc$^{2}$)$^{-1}$). The mangenta range
  shows estimates based on galaxy stellar mass functions using the
  dust-based interstellar medium mass scaling method as described by
  \cite{scoville17}. None of the measurements are extrapolated to
  account for the faint end of the molecular gas mass function that
  remained inaccessible to each survey. The red bar indicates the
  constraint obtained from intensity mapping by \cite{keating16}. No
  uncertainties are shown for this measurement, since they are
  dominated by model assumptions rather than statistical measurement
  errors. {\em Right:} Same data, but also showing the total
  star-formation rate density, multiplied by equivalent gas depletion
  timescales of 0.1, 0.5, and 1.0\,Gyr, for reference. Lighter shaded
  regions correspond to star-formation rate estimates based on
  ultraviolet stellar light measurements, with ``corrections'' for
  estimated losses due to dust extinction of the ultraviolet light
  applied. Darker shaded regions correspond to star-formation rate
  estimates based on direct measurements of the dust-obscured stellar
  light at infrared wavelengths (\citealt{bouwens16}, including
  infrared-bright sources from \citealt{magnelli13}; see, e.g.,
  \citealt{md14} for further details on uncertainties of the
  star-formation rate density measurements).
  \label{f5}}
\end{figure*}

\begin{table*}

\caption{Cold gas density evolution measurements from COLDz.\label{t3}}
\centering
\begin{tabular}{ c c c c }
\hline
  Redshift range & Lower limit & Median & Upper limit\\
  &(5$^{\rm th}$ percentile)&(50$^{\rm th}$ percentile) & (95$^{\rm th}$ percentile) \\
  &$10^7\,M_\odot\,$Mpc$^{-3}$&$10^7\,M_\odot\,$Mpc$^{-3}$&$10^7\,M_\odot\,$Mpc$^{-3}$\\
   \hline
 1.95--2.85 & 1.1  & 2.7 & 5.6\\
 1.95--2.72 & 0.95$^{\rm a}$ & 3.5$^{\rm a}$ & 10.9$^{\rm a}$ \\
 {\em 2.03--2.85} & {\em 0.30}$^{\rm b}$ & {\em 1.9}$^{\rm b}$ & {\em 7.3}$^{\rm b}$ \\
 4.90--6.70 & 0.14 & 0.47 & 1.1\\
            &      &      & 4.0$^{\rm c}$\\
 \hline
\end{tabular}
\vspace{0.1mm}

$^{\rm a}$Measurement for the COSMOS field alone, after merging both purity methods.\\
$^{\rm b}$Measurement for the GOODS-North field alone, after merging both purity methods. These data alone do not fully sample the ``knee'' of the CO luminosity function.\\
$^{\rm c}$Less constraining upper limit obtained when making the (unlikely) assumption that all \aco\ candidates not yet independently confirmed could, in principle, be \bco\ emitters.

\end{table*}

\subsection{COLDz Cold Gas Density of the Universe}

By integrating the measurements and upper limits obtained on the
\aco\ and \bco\ luminosity functions across each of the full redshift
intervals, we obtain estimates of the total CO luminosity density per
unit volume. For the higher-redshift bin, we use the results from the
two methods described above as a direct measurement and as a
conservative upper limit, respectively. As done in previous work
(e.g., \citealt{walter14,decarli16a}), we do not extrapolate the faint
end of the luminosity function, but instead only include measurements
down to the limit of our survey of log($L'_{\rm CO}$/\lprime
)$\simeq$9.5. Given the consistency of the COLDz data with a flat
faint-end slope, and the moderate survey statistics, this assumption
is not likely to dominate the uncertainty budget of our measurement,
but more sensitive observations are required to fully assess the
impact of this assumption.\footnote{For reference, integrating our
  best-fit model Schechter functions down only to log($L'_{\rm
    CO}$/\lprime )=10.0 would result in a 0.07\,dex lower median value
  for the integrated \aco\ luminosity density. Integrating down
  further to log($L'_{\rm CO}$/\lprime )=8.0 or 9.0 would result in
  0.044\,dex or 0.03\,dex higher values, respectively.} We then
convert the measurements of the CO luminosity density to a molecular
gas mass density by applying a ``standard'' conversion factor of
$\alpha_{\rm CO}$=3.6\,\msol\,(K\,\kms\,pc$^{2}$)$^{-1}$ (e.g.,
\citealt{daddi10a}). This choice is motivated by the finding that the
majority of the independently-confirmed \aco\ emitters (with the
exception of the major merger GN19) are consistent with the
star-forming galaxy ``main sequence'' at $z$$\sim$2--3. We do not
apply a separate correction to $\alpha_{\rm CO}$ for the \bco-based
measurement, since typical \bco/\aco\ line brightness temperature
ratios for ``normal'' high-redshift galaxies are of order 90\% (e.g.,
\citealt{cw13}), and since the actual \bco\ detections in our survey
are all massive dust-obscured starburst galaxies. As such, the implied
$\sim$10\% correction required is likely sub-dominant to the
assumptions made for the choice of $\alpha_{\rm CO}$ (which also
depends on other factors like metallicity; see \citealt{bolatto13} for
a review). The choice of $\alpha_{\rm CO}$ will be re-evaluated in
upcoming work, once dynamical mass estimates based on
spatially-resolved measurements of individual line candidates are
available, but we note that the choice of a smaller,
``starburst-like'' $\alpha_{\rm CO}$ of order unity would result in
significantly lower cold gas density estimates. The resulting
measurements of the cold gas density of the universe are shown in
Fig.~\ref{f5} and summarized in Table~\ref{t3}.

\section{Discussion}

Due to improved statistics, the COLDz data provide the currently best
constraints on the CO luminosity function at $z$$\sim$2--3 and
$z$$\sim$5--7, and they allow for a measurement of the shape of the CO
luminosity function in the $z$$\sim$2--3 bin. This provides the to
date perhaps most solid constraints on the cosmic density of cold
molecular gas in galaxies at these redshifts.

\subsection{Comparison to Previous ``Blind'' CO Surveys}

\subsubsection{CO Luminosity Function}

The most similar measurements of the CO luminosity function to COLDz
are those in the {\em Hubble} Deep Field North (HDF-N) and in the {\em
  Hubble} Ultra Deep Field (H-UDF; ASPECS-Pilot survey) over $\sim$0.5
and 1\,arcmin$^2$ size regions, covering the \cco\ line at
$\langle$$z$$\rangle$=2.75 and $\langle$$z$$\rangle$=2.61,
respectively (\citealt{walter14,walter16,decarli14,decarli16a}). We
consider the differences in redshift in these previous works to the
$\sim$60\,arcmin$^2$ COLDz \aco\ survey ($\langle$$z$$\rangle$=2.35
and 2.44 in the COSMOS and GOODS-North fields, respectively) presented
here negligible compared to other sources of uncertainty, such that we
directly compare these measurements in the following. The difference
in line search methods and the luminosity function calculation yield
perhaps more conservative uncertainty estimates for the COLDz
constraints, but we have confirmed that we would obtain consistent
results when adopting the same methods employed in the analysis of the
ASPECS-Pilot survey (\citealt{decarli16a}). We thus adopt the
measurements and uncertainties from the previous surveys without
further modifications.

As shown in Figures~\ref{f2} and \ref{f3}, we find that the
measurements of all three surveys are consistent within the relative
uncertainties. There may be tentative evidence that the COLDz
measurements are somewhat lower than the ASPECS-Pilot measurements in
the best-constrained common luminosity range at log($L'_{\rm
  CO}$/\lprime )$\simeq$10.2--11.0 (Figs.~\ref{f2} and \ref{f3}). If
real, this effect may be due to cosmic variance, or it could be an
indication that \cco-based surveys preferentially select galaxies with
higher gas excitation, such that \cco/\aco\ brightness temperature
ratio of $r_{31}$=0.42$\pm$0.07 assumed by \cite{decarli16a} to
correct for the average gas excitation may be too low (which could
then mimick such an effect in principle, depending on the intrinsic
shape of the CO luminosity function).\footnote{For a sample of bright,
  observed-frame 850\,$\mu$m--selected galaxies, \cite{bothwell13}
  find a median $r_{31}$ of 0.52$\pm$0.09, but while there is source
  overlap, these galaxies are typically more intensely star-forming
  than the majority of sources found in the ``blind'' CO
  surveys. Also, the $\sim$20\%$\pm$20\% difference in $r_{31}$ is
  perhaps not sufficient to fully explain the observed effect.} The
latter would be consistent with the finding of a high line ratio limit
of $r_{31}$$>$0.7 for a candidate overlapping between the HDF-N and
COLDz surveys (ID19; \citealt{decarli14}), and the lack of
\aco\ detections for other mid-$J$ CO candidates in the same field
(see Paper I). Since some of these earlier candidates may be spurious,
and given the limited statistics of the current surveys and the
limited magnitude of the effect, additional data are required to
further investigate the relevance of potential selection effects due
to CO excitation. In particular, \cite{decarli16b} find that some
confirmed sources in the ASPECS-Pilot survey appear to show
comparatively low CO excitation, opposite to what would be expected in
the case of a CO excitation-based selection bias. The full, extended
ASPECS survey data expected from an ongoing ALMA Large Program will
further constrain the contribution of cosmic variance to the observed
effect.

\subsubsection{Cold Gas Density of the Universe}

The constraints on the evolution of the cold gas density with redshift
resulting from the improved CO luminosity function measurements
provided by COLDz are consistent with those from previous surveys
within the relative uncertainties, and they extend the range of
estimates to earlier cosmic epochs (Fig.~\ref{f5}; all CO surveys
assume the same $\alpha_{\rm CO}$). As in the case of the CO
luminosity function constraints, we adopt the measurements and
uncertainties from previous works without further modifications. Due
to differences in the methods used to determine and report
uncertainties, caution is advised when comparing the constraints from
different surveys at face value.\footnote{For the HDF-N measurements,
  the lower and upper limits of the boxes shown represent secure
  detections and all line candidates reported by \cite{walter14},
  respectively, with Poissonian uncertainties due to the number of
  candidates added as error bars. For the ASPECS-Pilot measurements,
  box sizes indicate Poissonian errors, with a minor contribution due
  to flux errors and potential line misidentifications added
  (\citealt{decarli16a}). Both surveys adopt 1$\sigma$ uncertainty
  ranges, rather than the more conservative 90\% confidence intervals
  adopted for the COLDz measurements.} Previous surveys carried out at
3\,mm and 1\,mm did not provide estimates at $z$$>$4.5, since those
redshifts are only covered in high-$J$ lines, where estimates of CO
excitation as necessary to extrapolate the \aco\ luminosity are
increasingly uncertain.

The COLDz measurements likely suggest a higher gas density at
$z$$\sim$2--3
($\rho$(H$_2$)=0.95--10.9$\times$10$^7$\,\msol\,Mpc$^{-3}$, with a
preferred range of
1.1--5.6$\times$10$^7$\,\msol\,Mpc$^{-3}$)\footnote{For reference, the
  contribution from independently confirmed sources alone is
  $\rho$(H$_2$)=(2.8$\pm$1.4)$\times$10$^7$\,\msol\,Mpc$^{-3}$, which
  is consistent with the median value of the total of
  2.7$\times$10$^7$\,\msol\,Mpc$^{-3}$. Paper I also reports a weak
  \aco\ detection from stacking 34 individually-undetected
  $z$=2.0--2.8 galaxies with stellar masses of
  $M_{\star}$$>$10$^{10}$\,\msol\ in the GOODS-North field. As they
  are not detected individually, these galaxies are not part of the
  statistical sample used in this paper. If we were to include these
  sources, they would contribute an additional
  $\rho$(H$_2$)$\simeq$0.3$\times$10$^7$\,\msol\,Mpc$^{-3}$ in
  aggregate.} compared to $z$=0
($\rho$(H$_2$)=1.1$^{+0.7}_{-0.5}$$\times$10$^7$\,\msol\,Mpc$^{-3}$;
\citealt{saintonge17}; see also
\citealt{keres03,boselli14})\footnote{We here and in Fig.~\ref{f5}
  adopt $\alpha_{\rm CO}$=3.6\,\msol\,(K\,\kms\,pc$^{2}$)$^{-1}$ for
  consistency. Adopting $\alpha_{\rm
    CO}$=6.5\,\msol\,(K\,\kms\,pc$^{2}$)$^{-1}$ as used by
  \cite{keres03} would result in a factor of 1.8 higher
  $\rho$(H$_2$).}  by a factor of a few. This finding is consistent
with what was reported by the ASPECS team within the relative
uncertainties ($\rho$(H$_2$)=4.9--19$\times$10$^7$\,\msol\,Mpc$^{-3}$;
\citealt{decarli16a}). These measurements are also in agreement with
estimates based on galaxy stellar mass functions in COSMOS using the
dust-based interstellar medium mass scaling method as described by
\cite{scoville17}. Averaging their data at $z$=2.25 and 2.75, Scoville
et al.\ suggest $\rho$(ISM)=3.8$\times$10$^7$\,\msol\,Mpc$^{-3}$ at
$z$=2.5.\footnote{We here and in Fig.~\ref{f5} adopt $\alpha_{\rm
    CO}$=3.6\,\msol\,(K\,\kms\,pc$^{2}$)$^{-1}$ for
  consistency. Adopting $\alpha_{\rm
    CO}$=6.5\,\msol\,(K\,\kms\,pc$^{2}$)$^{-1}$ as used by
  \cite{scoville17} would result in a factor of 1.8 higher
  $\rho$(H$_2$).}  The latter agrees to within $\sim$30\% with the
median value of the COLDz measurement, and within $<$10\% with the
median value measured in the COSMOS field alone. The COLDz results are
also consistent with the constraints obtained from \aco\ intensity
mapping experiments at similar redshifts in the GOODS-North field
($\rho$(H$_2$)=9.2$^{+5.9}_{-3.3}$$\times$10$^7$\,\msol\,Mpc$^{-3}$ at
$z$=2.3--3.3; \citealt{keating16}).\footnote{We here and in
  Fig.~\ref{f5} adopt $\alpha_{\rm
    CO}$=3.6\,\msol\,(K\,\kms\,pc$^{2}$)$^{-1}$ for
  consistency. Adopting $\alpha_{\rm
    CO}$=4.3\,\msol\,(K\,\kms\,pc$^{2}$)$^{-1}$ as used by
  \cite{keating16} would result in a factor of 1.2 higher
  $\rho$(H$_2$).} A comparison of these results is valuable in
general, since CO intensity maps in principle may contain signal below
the detection threshold of galaxy surveys, but we note that a
quantitative comparison of the relative uncertainties is
difficult. This is due to the fact that the intensity mapping
constraints only measure the second raw moment of the luminosity
function, and therefore cannot distinguish between contributions due
to the characteristic luminosity and volume density to the
measurement. Furthermore, the detailed interpretation of the nature of
the intensity mapping signal in principle relies on assuming a scaling
relation between dark matter halo mass and CO luminosity, which is
currently not well constrained at $z$$\sim$2--3. We thus do not show
formal error bars for this measurement in Fig.~\ref{f5}.

The COLDz measurements are also consistent with a decrease in gas
density from $z$$\sim$2--3 towards $z$$\sim$5--7
($\rho$(H$_2$)=0.14--1.1$\times$10$^7$\,\msol\,Mpc$^{-3}$), possibly
to below the present-day value. The redshift evolution of the cold gas
history of the universe thus appears qualitatively similar to that of
the star-formation history of the universe (e.g., \citealt{md14}),
which is consistent with what is expected if a universal
``star-formation law'' between gas mass and star-formation rate (e.g.,
\citealt{cw13}) already exists at early epochs.

\subsubsection{Gas Depletion Times}

In combination, the cold gas mass and star-formation rate density
evolution entail information about the evolution of galaxy gas
depletion times as a function of redshift.\footnote{The analysis
  presented here concerns the redshift evolution of gas depletion
  times, and thus, does not further consider the potential range of
  values expected for different galaxy populations that contribute
  to the signal.} As shown in Fig.~\ref{f5}, simply multiplying the
total star-formation rate density (\citealt{bouwens16}) by a
characteristic gas depletion timescale (which, to first order,
represents the ratio between molecular gas mass and star-formation
rate, $M_{\rm H_2}$/SFR) of several hundred million years provides a
reasonable match to the cold gas density relation at all redshifts
currently probed within the uncertainties, although the data may
tentatively prefer shorter depletion times towards higher
redshifts. At $z$=0, a characteristic gas depletion timescale in the
range of $\tau_{\rm dep}^{\rm ch}$=0.5--1\,Gyr is preferred with the
adopted $\alpha_{\rm CO}$ conversion factor (see also discussion by
\citealt{saintonge17}).  Based on the COLDz measurements of
$\rho$(H$_2$) at $z$=2.4 and adopting
$\rho$(SFR)=0.15$\pm$0.05\,\msol\,yr$^{-1}$\,Mpc$^{-3}$ (e.g.,
\citealt{md14}), we find a characteristic gas depletion timescale of
$\tau_{\rm dep}^{\rm ch}$=70--750\,Myr with 90\% confidence, with a
median value of 200$\pm$70\,Myr. Since the star-formation rate density
relation includes significantly less luminous galaxies than probed by
current blind CO surveys, this may either indicate that low-luminosity
galaxies below our $L'_{\rm CO}$ detection limit do not contribute
dominantly to the total cold gas density (perhaps implying that the
faint-end slope of the CO luminosity function is not steeply rising
towards lower $L'_{\rm CO}$), or that the characteristic gas depletion
timescales are longer than 200--500\,Myr when averaged over the entire
galaxy population. Assuming substantially shorter gas depletion
timescales (or, high star-formation efficiencies) appears to be
inconsistent with the data, unless the characteristic $\alpha_{\rm
  CO}$ conversion factor is substantially lower than assumed. The
COLDz measurements of $\rho$(H$_2$) at $z$=5.8 are consistent with
characteristic gas depletion timescales of $\tau_{\rm dep}^{\rm
  ch}$$>$100\,Myr, with a factor of a few higher values allowed by the
data within the uncertainties. Although not a unique conclusion based
on the COLDz data given the remaining uncertainties, a shortening in
gas depletion times despite the observed increase in cold molecular
gas content in star-forming galaxies towards higher redshift would be
consistent with similar findings based on targeted studies of
\cco\ emission and dust-based interstellar medium mass estimates
(e.g., \citealt{genzel15,scoville17}), and thus, with an effective
increase in star formation efficiency (i.e., SFR per unit $M_{\rm
  H_2}$) towards higher redshifts.

\subsection{Comparison to Model Predictions}

\subsubsection{CO Luminosity Function}

Given the consistency between the COLDz data and previous surveys, we
compare the new CO luminosity function measurements to predictions
based on semi-analytical models (\citealt{lagos12,popping16}) and
empirical estimates based on the infrared luminosity function of {\em
  Herschel}-selected galaxies under the assumption of a
``star-formation law'' (\citealt{vallini16}; Figs.~\ref{f2} to
\ref{f4}; see, e.g., \citealt{lagos15,dave17,xie17} for additional
model predictions).

The measurements at $z$$\sim$2--3 appear to be inconsistent with the
semi-analytical predictions (see \citealt{decarli16a} for a detailed
comparison of both models), which place the characteristic luminosity
$L'^*_{\rm CO}$ (``knee'') of the luminosity function at significantly
lower luminosities than observed. This is consistent with the excess
of bright sources compared to the predictions seen in the ASPECS-Pilot
data alone in some luminosity bins (\citealt{decarli16a}), but the
trend becomes clearer at the higher statistical significance of the
COLDz measurements -- showing a significantly (by one to two orders of
magnitude) higher characteristic luminosity than what is observed at
$z$$\sim$0 (e.g., \citealt{keres03,boselli14,saintonge17}). These
predictions also prefer a strong contribution from faint sources,
which are not preferred by the data, but agree within the considerable
uncertainties of the measurements at low luminosities. Qualitatively,
the underprediction in the number of luminous CO emitters may be
related to the finding that semi-analytical models tend to
underpredict the star-formation rates of galaxies on the star-forming
main sequence at similar redshifts (see, e.g., review by
\citealt{sd15}).

At $z$$\sim$5--7, the excess of bright sources compared to the
semi-analytical predictions appears to be even more pronounced than at
lower redshifts, but we caution that the most constraining measurement
is based on a small number of sources only, and thus, needs to be put
on a firmer statistical footing. On the other hand, the observations
at $z$$\sim$2--3 appear to be consistent with the empirical
predictions by \cite{vallini16}, and thus, with what is expected from
estimates of dust-obscured star-formation activity at high redshift
based on infrared luminosity functions.

\subsubsection{Cold Gas Density of the Universe}

The observational constraints on the evolution of the cold gas density
with redshift remain in agreement with both semi-analytical
(\citealt{obreschkow09,lagos11,lagos12,popping14b,popping14a}) and
empirical (\citealt{sargent12,sargent14}) model predictions at
$z$$\sim$2--3. Since most of the semi-analytical models include a
varying $\alpha_{\rm CO}$ between individual galaxies, a simple
interpretation of the consistency despite the disagreement in the
luminosity function estimates remains challenging. In addition, the
predictions do not account for the sensitivity limits of the CO
surveys, or uncertainties due to cosmic variance. Given the
differences in the CO luminosity functions between models and
observations at high $z$, this effect could lead to up to a factor of
a few difference in the corresponding gas densities at high redshift
in principle. However, as discussed in Sect.~3.4, the impact of the
sensitivity limits, which would bias the measurements towards lower
values, appears to be relatively minor based on the preferred model
Schechter function fits to the COLDz data. On the other hand, cosmic
variance due to large-scale structure in the distribution of gas-rich
galaxies (which is assumed to be uniform in our analysis) could bias
the measurements either low or high.
To obtain an approximate estimate of systematic uncertainties
introduced by cosmic variance, we follow the prescription by
\cite{driver10}, based on the distribution of galaxies near the
characteristic stellar mass at a given redshift.\footnote{This method
  is based on a generalized expression (their equation 4) calibrated
  through an examination of galaxies within $\pm$1\,mag of the
  characteristic optical magnitude out to $z$$\sim$0.1 in the Sloan
  Digital Sky Survey (SDSS), which represent the most common galaxies
  at a given redshift. See {\tt cosmocalc.icrar.org} for additional
  details. The field sizes and \aco\ redshift ranges in Table \ref{t1}
  are used for all calculations.} Including both Poisson uncertainty
and cosmic variance scaled to the volume of the COLDz survey, we find
a sample variance uncertainty of $\sim$30\%--40\% and $\sim$25\% for
the COSMOS and GOODS-North fields, respectively. Alternatively,
adapting results based on models of the evolution of the most massive
galaxies (i.e., $M_\star$$>$10$^{11}$\,$M_\odot$) by
\cite{moster11}\footnote{For our estimate, we have made use of the
  predictions provided for the H-UDF, since this field has a similar
  area as the COLDz COSMOS field. We have assumed $\langle z
  \rangle$=2.35 and $\Delta$$z$=0.8 in our calculations, and we adopt
  the values found for the
  $M_\star$=10$^{11.0}$--10$^{11.5}$\,$M_\odot$ bin.} yields an
estimated uncertainty of $\sim$40\%--50\% based on cosmic variance
alone for the smaller COSMOS field. These estimates are consistent
with what is found from more detailed calculations based on the {\em
  IllustrisTNG} simulations (of order $\sim$30\%--50\% in the
$z$$\sim$2--3 bin for both fields combined; G.\ Popping 2018, private
communication). All estimates appear to suggest that uncertainties due
to cosmic variance are subdominant to other sources of uncertainty,
given the large volume of the COLDz survey due to the broad range in
redshift covered and the comparatively large field size. This is also
consistent with the broad distribution in redshift of the confirmed CO
emitters and candidates in the COLDz survey volume (see, e.g., Paper
I, Fig.~3).\footnote{The COSMOS field contains the AzTEC-3
  protocluster region at $z$=5.3 (e.g.,
  \citealt{capak11,riechers10a,riechers14b}), and thus, is biased in
  principle. However, only upper limits are reported in this field for
  the corresponding redshift bin, such that this does not impact the
  reported measurements.} In any case, if we were to conservatively
correct down the model predictions by factors of $\sim$1.5--2 to
account for the combined effects of sensitivity limits and cosmic
variance, they would in fact move close to the median $\rho$(H$_2$)
implied by the COLDz measurements. Taken at face value, the apparent
agreement between the model predictions and COLDz data could indicate
that there is no significant, or at least, no dominant contribution
from sources far below the COLDz detection limit, such that
steeply-rising faint-end slopes of the CO luminosity function towards
lower $L'_{\rm CO}$ may be disfavored. This would also be consistent
with the agreement between the COLDz measurement and intensity mapping
constraints.

If true, this could be related to lower metallicities towards fainter,
low-mass galaxies, leading to disproportionally low CO luminosity per
unit molecular gas mass (e.g., \citealt{genzel12,bolatto13}). Although
not a unique explanation, this would be consistent with the finding of
a lower median redshift of galaxies with low submillimeter continuum
fluxes compared to brighter ones (implying low dust masses, and thus,
likely low gas masses; \citealt{aravena16}), and with the apparent
finding of a low dust content in lower stellar mass galaxies at
$z$$>$2 (\citealt{bouwens16}). This would also be consistent with the
finding that we do not detect \bco\ emission from several known,
modestly massive and star-forming Lyman-break galaxies at
$z$$\sim$5.2--5.3 in our survey area (Paper I; see also
\citealt{capak11,walter12,riechers14b}), which is compatible with a
perhaps elevated $\alpha_{\rm CO}$ due to lower metallicity.  All
confirmed $z$$>$5 COLDz detections are massive, dust-obscured
starburst galaxies with likely high metallicity.

The observational constraints at $z$$\sim$5--7 are also in agreement
with the model predictions, albeit lower than the Obreschkow et
al.\ and Lagos et al.\ models unless some unconfirmed sources (which
are taken into consideration for the upper limit shown) contribute to
the signal.  They are less secure than in the lower-redshift bin due
to more limited statistics and because it is currently not possible to
measure the characteristic CO luminosity at these redshifts, such that
the fraction of the total cold gas density recovered down to the
sensitivity limit of the survey is less certain than at lower
redshifts. Nonetheless, this finding appears to be consistent with the
assumption of an evolving $\alpha_{\rm CO}$ due to lower metallicity
in fainter galaxies and towards higher redshifts, resulting in a steep
drop in the gas volume density as traced by CO emission. Further
observations are required to investigate if the drop in H$_2$ density
towards very high redshift is as steep as observed in CO, or if the
effect is enhanced due to metallicity affecting the strength of the CO
signal.\footnote{The strength of the CO signal may also be reduced at
  the highest redshifts due to the increased temperature of the cosmic
  microwave background, relative to which the cold gas emission is
  detected. The importance of this effect strongly depends on the
  excitation of the gas traced by CO, in particular the kinetic gas
  temperature (e.g., \citealt{dacunha13}).}

\section{Conclusions}

We have used the ``blind'' molecular line scans over
$\sim$60\,arcmin$^2$ in the COSMOS and GOODS-North survey fields taken
as part of the VLA COLDz survey (Paper I) to measure the shape of the
CO luminosity function at $z$$\sim$2--3 and to constrain it at
$z$$\sim$5--7, utilizing \aco\ and \bco\ emission line galaxy
candidates. We also provide constraints on the evolution of the cosmic
molecular gas density out to $z$$\sim$7. We compare our findings to
previous $\sim$0.5 and 1\,arcmin$^2$ surveys in the HDF-N and the
H-UDF (ASPECS-Pilot) in higher-$J$ CO lines
(\citealt{decarli14,walter16}), estimates based on galaxy stellar mass
functions in COSMOS scaled using dust-based interstellar medium mass
estimates (\citealt{scoville17}), and a CO intensity mapping study in
GOODS-North (\citealt{keating16}), finding broad agreement within the
relative uncertainties. The COLDz data provide the first solid
measurement of the shape of the CO luminosity function at
$z$$\sim$2--3, reaching below its ``knee'', and the first significant
constraints at $z$$\sim$5--7. The characteristic CO luminosity at
$z$$\sim$2--3 appears to be one to two orders of magnitude higher than
at $z$=0 (\citealt{keres03,saintonge17}), which is consistent with the
idea that the dominant star-forming galaxy populations
$\sim$10\,billion years ago were significantly more gas-rich compared
to present day. We also independently confirm an observed apparent
excess of the space density of bright CO-emitting sources at high
redshift compared to semi-analytical predictions, but our findings are
consistent with empirical predictions based on the infrared luminosity
function and observed star-formation rates of distant galaxies.

Integrating the CO luminosity functions down to the sensitivity limit
of our survey, we obtain robust estimates of the volume density of
cold gas in galaxies at high redshift. Our measurement is consistent
with a factor of a few increase from $z$$\sim$0 to $z$$\sim$2--3, and
a decrease towards $z$$\sim$5--7 by about an order of magnitude (which
may be less steep in practice if metallicity has an increasing effect
on CO-based measurements towards the highest redshifts). This is
consistent with semi-analytical and empirical model predictions and
previous constraints from the ASPECS-Pilot survey
(\citealt{decarli16a}), and with previous findings of increased gas
fractions at $z$$>$1--2 (e.g.,
\citealt{daddi10a,tacconi13,tacconi18,scoville17}). The overall shape
of the cosmic gas density evolution resembles that of the
star-formation history of the universe, consistent with an underlying
``star-formation law'' relation out to the highest measured
redshifts. This suggests that the star-formation history, to first
order, follows the evolution of the molecular gas supply in galaxies,
as regulated by the gas accretion efficiency and feedback processes. A
more direct comparison of the star-formation rate and cold gas density
relations as a function of cosmic time holds critical information
about the true gas depletion timescales, and thus, the gas accretion
rates required to maintain the ongoing build-up of stellar mass. The
data appear broadly consistent with a characteristic gas depletion
timescale of several hundred million years, but there may be tentative
evidence for a shortening in gas depletion times despite the observed
increase in cold molecular gas content in star-forming galaxies
towards higher redshift. This finding would be consistent with
previous, targeted investigations based on \cco\ and dust-based
interstellar medium mass estimates (e.g.,
\citealt{genzel15,scoville17}), and thus, with an effective increase
in star formation efficiency in the dominant star-forming galaxy
populations towards higher redshifts.

While COLDz is the currently largest survey of its kind, the size of
the volume probed and the number of line candidates found implies that
larger areas need to be surveyed to greater depth in the future to
more clearly address effects of cosmic variance and to reduce the
error budget due to Poissonian fluctuations. Such studies will be
possible with large investments of observing time at the VLA and ALMA
in the coming years, until the next large leap in capabilities will
become available with the construction of the Next Generation Very
Large Array (ngVLA; e.g., \citealt{bolatto17,selina18}).

\acknowledgments

The authors thank Am\'elie Saintonge for sharing her updated Schechter
fits to the local CO luminosity function prior to publication, Rychard
Bouwens for sharing the data necessary to create Figure \ref{f5}, and
Gerg\"o Popping for helpful input on the comparison of our results to
models, and for providing some new and updated model results. We also
thank the anonymous referee for a helpful report. D.R.\ and
R.P.\ acknowledge support from the National Science Foundation under
grant number AST-1614213 to Cornell University. J.A.H. acknowledges
support of the VIDI research program with project number 639.042.611,
which is (partly) financed by the Netherlands Organisation for
Scientific Research (NWO). I.R.S. acknowledges support from the ERC
Advanced Grant DUSTYGAL (321334), STFC (ST/P000541/1), and a Royal
Society/Wolfson Merit Award. The National Radio Astronomy Observatory
is a facility of the National Science Foundation operated under
cooperative agreement by Associated Universities, Inc.

\vspace{5mm}
\facilities{VLA}

\appendix

\section{Luminosity Function Constraints:\ Tabulated Results}

For reference, we here include the measured ranges of the CO
luminosity function from the COLDz \aco\ data at
$\langle$$z$$\rangle$=2.4 (Table~\ref{ta1}) and the \bco\ data at
$\langle$$z$$\rangle$=5.8 (Tables~\ref{ta2} and \ref{ta3}), as
utilized in Figures~\ref{f2} to \ref{f4}. log($L'_{\rm CO}$) bins are
0.5\,dex wide and given in steps of 0.1\,dex, such that every 5$^{\rm
  th}$ bin is statistically independent.

\begin{table*}[h]
{  \centering
\caption{Measured ranges of the \aco\ luminosity function at $z$$\sim$2.4 from the COLDz data (5$^{\rm th}$ and 95$^{\rm th}$ percentiles).}
\label{ta1}
\begin{tabular}{ c | c  c | c c  | c }
\hline\hline
log($L'_{\rm CO}$) bin & COSMOS ``uniform''$^{\rm a}$ & COSMOS ``normal''$^{\rm a}$ &  GOODS-N ``uniform''$^{\rm a}$ & GOODS-N ``normal''$^{\rm a}$ & Combined Fields ``merged''$^{\rm a}$\\
$$ [K\,\kms\,pc$^2$] & [Mpc$^{-3}$\,dex$^{-1}$] & [Mpc$^{-3}$\,dex$^{-1}$] & [Mpc$^{-3}$\,dex$^{-1}$] & [Mpc$^{-3}$\,dex$^{-1}$] & [Mpc$^{-3}$\,dex$^{-1}$] \\
\hline
{\bf 9.5 -- 10.0} & --4.04, --2.55& --3.77, --2.21&  $\cdots$&  $\cdots$& --4.04, --2.21\\
9.6 -- 10.1 & --4.08, --3.03& --3.83, --2.69&  $\cdots$&  $\cdots$& --4.08, --2.69\\
9.7 -- 10.2 & --4.14, --3.32& --3.87, --2.95&  $\cdots$&  $\cdots$& --4.14, --2.95\\
9.8 -- 10.3 & --4.19, --3.47& --3.92, --3.12&  $\cdots$&  $\cdots$& --4.19, --3.12\\
9.9 -- 10.4 & --4.23, --3.44& --3.96, --3.24&  $\cdots$&  $\cdots$& --4.23, --3.24\\
{\bf 10.0 -- 10.5} & --4.26, --3.38& --3.98, --3.26& --4.25, --3.37& --3.97, --3.00& --4.10, --3.32\\
10.1 -- 10.6 & --4.16, --3.41& --3.81, --3.27& --4.27, --3.61& --3.97, --3.27& --4.07, --3.41\\
10.2 -- 10.7 & --3.70, --3.31& --3.58, --3.26& --4.32, --3.79& --4.03, --3.47& --3.76, --3.37\\
10.3 -- 10.8 & --3.62, --3.36& --3.60, --3.29& --4.39, --3.92& --4.11, --3.62& --3.69, --3.39\\
10.4 -- 10.9 & --3.65, --3.41& --3.63, --3.34& --4.47, --4.01& --4.20, --3.74& --3.73, --3.45\\
{\bf 10.5 -- 11.0} & --3.85, --3.45& --3.77, --3.40& --4.59, --4.11& --4.31, --3.86& --3.98, --3.60\\
10.6 -- 11.1 & --3.96, --3.49& --3.96, --3.48& --4.71, --4.21& --4.43, --3.97& --4.10, --3.70\\
10.7 -- 11.2 & --5.23, --3.69& --4.90, --3.69& --4.82, --4.31& --4.56, --4.07& --4.79, --4.11\\
10.8 -- 11.3 & $<$(--8.00), --3.95& $<$(--8.00), --3.90& --4.96, --4.41& --4.73, --4.19& --5.00, --4.25\\
10.9 -- 11.4 & $<$(--8.00), --4.00& $<$(--8.00), --4.00& --5.38, --4.51& --5.13, --4.32& --5.44, --4.41\\
{\bf 11.0 -- 11.5} &  $\cdots$&  $\cdots$& --6.28, --4.62& --6.23, --4.45& --6.28, --4.45\\
11.1 -- 11.6 &  $\cdots$&  $\cdots$& $<$(--8.00), --4.72& $<$(--8.00), --4.61& $<$(--8.00), --4.61\\
11.2 -- 11.7 &  $\cdots$&  $\cdots$& $<$(--8.00), --4.80& $<$(--8.00), --4.77& $<$(--8.00), --4.77\\
11.3 -- 11.8 &  $\cdots$&  $\cdots$& $<$(--8.00), --4.82& $<$(--8.00), --4.82& $<$(--8.00), --4.82\\
11.4 -- 11.9 &  $\cdots$&  $\cdots$& $<$(--8.00), --5.96& $<$(--8.00), --5.56& $<$(--8.00), --5.56\\

\hline \noalign {\smallskip}
\end{tabular}
}
\vspace{0.1mm}

$^{\rm a}$Given as log($\Phi_{\rm CO})$. Based on ``uniform'', ``normal'', and merged (last column) purity uncertainty estimates as described in Section 3. 

\end{table*}

\begin{table}[t]
{  \centering
\caption{Measured ranges of the \bco\ luminosity function at $z$$\sim$5.8 from the COLDz data (90$^{\rm th}$ percentile upper limits).}
\label{ta2}
\begin{tabular}{ c | c | c }
\hline\hline 
log($L'_{\rm CO}$) bin & COSMOS$^{\rm a}$ &  GOODS-N$^{\rm a}$ \\
 $$[K$\,$\kms\,pc$^2$] & [Mpc$^{-3}\,$dex$^{-1}$] & [Mpc$^{-3}\,$dex$^{-1}$]  \\
\hline
{\bf 9.5 -- 10.0}  & $<$(--2.92) & $\cdots$\\
9.6 -- 10.1  & $<$(--2.94) & $\cdots$\\
9.7 -- 10.2  & $<$(--3.20) & $\cdots$\\
9.8 -- 10.3  & $<$(--3.42) & $\cdots$\\
9.9 -- 10.4  & $<$(--3.57) & $\cdots$\\
{\bf 10.0 -- 10.5} & $<$(--3.64) & $<$(--3.10)\\
10.1 -- 10.6 & $<$(--3.68) & $<$(--3.41)\\
10.2 -- 10.7 & $<$(--3.74) & $<$(--3.63)\\
10.3 -- 10.8 & $<$(--3.81) & $<$(--3.80)\\
10.4 -- 10.9 & $<$(--3.88) & $<$(--3.94)\\
{\bf 10.5 -- 11.0} & $<$(--3.95) & $<$(--4.06)\\
10.6 -- 11.1 & $<$(--4.03) & $<$(--4.16)\\
10.7 -- 11.2 & $<$(--4.15) & $<$(--4.25)\\
10.8 -- 11.3 & $\cdots$& $<$(--4.34)\\
10.9 -- 11.4 & $\cdots$& $<$(--4.45)\\
{\bf 11.0 -- 11.5} & $\cdots$& $<$(--4.58)\\
11.1 -- 11.6 & $\cdots$& $<$(--4.73)\\
11.2 -- 11.7 & $\cdots$& $<$(--4.88)\\
11.3 -- 11.8 & $\cdots$& $<$(--5.24)\\
11.4 -- 11.9 & $\cdots$& $<$(--5.87)\\

\hline \noalign {\smallskip}
\end{tabular}
}
\vspace{0.1mm}

$^{\rm a}$Given as log($\Phi_{\rm CO})$. Based on ``normal'' purity uncertainty estimates as described in Section 3. 

\end{table}

\begin{table}[t]
{  \centering
\caption{\bco\ luminosity function at $z$$\sim$5.8 from the COLDz data based on confirmed sources only (red symbols in Fig.~\ref{f4}).}
\label{ta3}
\begin{tabular}{ c | c | c }
\hline\hline 
log($L'_{\rm CO}$) bin & COSMOS$^{\rm a}$ &  GOODS-N$^{\rm a}$ \\
 $$[K$\,$\kms\,pc$^2$] & [Mpc$^{-3}\,$dex$^{-1}$] & [Mpc$^{-3}\,$dex$^{-1}$]  \\
\hline
9.8 -- 10.3 & $<$(--3.82) & $\cdots$\\
10.7 -- 11.2 & $\cdots$& --5.07, --4.19\\

\hline \noalign {\smallskip}
\end{tabular}
}
\vspace{0.1mm}

$^{\rm a}$Given as log($\Phi_{\rm CO})$. See Section 3.3.2 for details. 

\end{table}

\section{Luminosity Function Modeling:\ Further Details}

\begin{figure*}
\epsscale{1.15}
\plotone{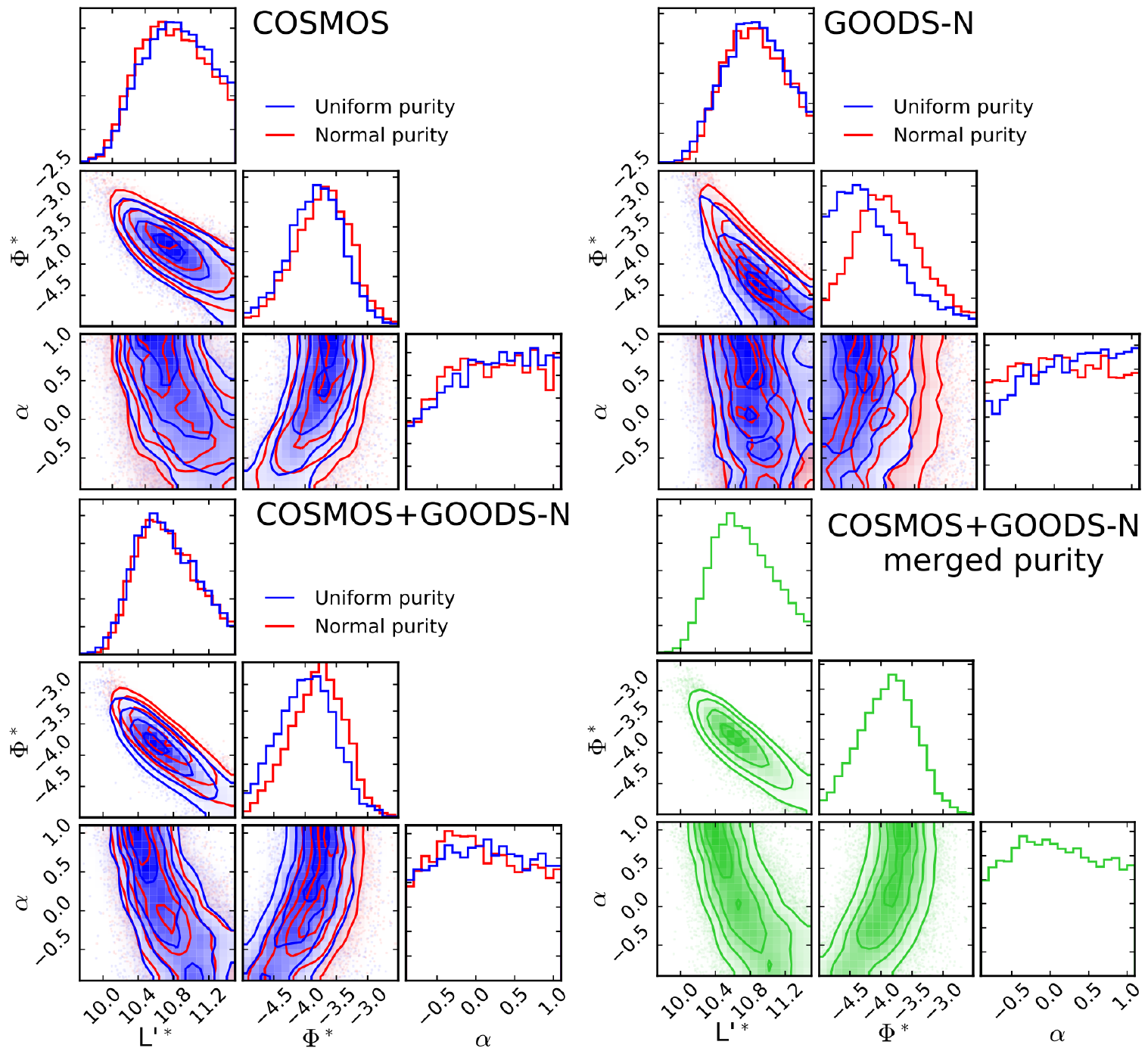}
\vspace{-2mm}

\caption{Corner plots of the Schechter model parameter posterior
  distribution from fitting the \aco\ luminosity function with the ABC
  method. Contour levels correspond to 0.5, 1.0, 1.5, and 2.0$\sigma$
  in two dimensions. {\em Top:} Parameters found when fitting results
  obtained with the ``uniform'' (blue) and ``normal'' (red) purity
  methods for the COLDz COSMOS ({\em left}) and GOODS-North ({\em
    right}) fields, respectively. {\em Bottom:} Same as found when
  combining the constraints from both fields, before ({\em left}) and
  after ({\em right}) merging the two methods used to calculate
  purities, respectively. \label{fb1}}
\end{figure*}

We here show the corner plots of the Schechter model parameter
posterior distribution from fitting the COLDz \aco\ luminosity
function with the ABC method (Figure~\ref{fb1}). The adopted prior
ranges are log($L'_{\rm CO}$/\lprime )=9.5 to 11.5, log($\Phi_{\rm
  CO}$/Mpc$^{-3}$\,dex$^{-1}$)=--5 to --2.5, and $\alpha$=--1 to 1.

\newpage

\bibliographystyle{yahapj}
\bibliography{ref.bib}


\end{document}